\documentclass[twocolumn,english,amsmath,amssymb,aps,prl,showpacs]{revtex4}
\usepackage[T1]{fontenc}
\usepackage[latin9]{inputenc}
\usepackage{color}
\usepackage{babel}
\usepackage{float}
\usepackage{amsmath}
\usepackage{amssymb}
\usepackage{wasysym}
\usepackage{graphicx}
\usepackage{esint}
\usepackage[unicode=true,
 bookmarks=false,
 breaklinks=false,pdfborder={0 0 1},backref=false,colorlinks=true]
 {hyperref}
\hypersetup{
 colorlinks,linkcolor=blue,citecolor=blue,urlcolor=blue}

\makeatletter

\providecommand{\tabularnewline}{\\}

\@ifundefined{textcolor}{}
{%
 \definecolor{BLACK}{gray}{0}
 \definecolor{WHITE}{gray}{1}
 \definecolor{RED}{rgb}{1,0,0}
 \definecolor{GREEN}{rgb}{0,1,0}
 \definecolor{BLUE}{rgb}{0,0,1}
 \definecolor{CYAN}{cmyk}{1,0,0,0}
 \definecolor{MAGENTA}{cmyk}{0,1,0,0}
 \definecolor{YELLOW}{cmyk}{0,0,1,0}
}

\usepackage{color}
\usepackage{babel}

\@ifundefined{textcolor}{}{%
 \definecolor{BLACK}{gray}{0}
 \definecolor{WHITE}{gray}{1}
 \definecolor{RED}{rgb}{1,0,0}
 \definecolor{GREEN}{rgb}{0,1,0}
 \definecolor{BLUE}{rgb}{0,0,1}
 \definecolor{CYAN}{cmyk}{1,0,0,0}
 \definecolor{MAGENTA}{cmyk}{0,1,0,0}
 \definecolor{YELLOW}{cmyk}{0,0,1,0}
}

\usepackage{color}
\usepackage{babel}

\@ifundefined{textcolor}{}{%
 \definecolor{BLACK}{gray}{0}
 \definecolor{WHITE}{gray}{1}
 \definecolor{RED}{rgb}{1,0,0}
 \definecolor{GREEN}{rgb}{0,1,0}
 \definecolor{BLUE}{rgb}{0,0,1}
 \definecolor{CYAN}{cmyk}{1,0,0,0}
 \definecolor{MAGENTA}{cmyk}{0,1,0,0}
 \definecolor{YELLOW}{cmyk}{0,0,1,0}
}

\usepackage{color}
\usepackage{babel}
\usepackage{babel}

\makeatother

\begin{document}

\title{Electrically tunable resonant scattering in fluorinated bilayer graphene}

\author{Adam A. Stabile$^{1}$}

\author{Aires Ferreira$^{2,3}$}

\email{aires.ferreira@york.ac.uk}

\selectlanguage{english}%

\author{Jing Li$^{1}$}

\author{N. M. R. Peres$^{3}$}

\author{J. Zhu$^{1,4}$}

\email{jzhu@phys.psu.edu}

\selectlanguage{english}%

\affiliation{$^{1}$Department of Physics, The Pennsylvania State University,
University Park, Pennsylvania 16802, USA}

\affiliation{$^{2}$Department of Physics, University of York, York YO10 5DD,
United Kingdom}

\affiliation{$^{3}$Centro de Fisica and Departamento de Fisica, Universidade
do Minho, Campus de Gualtar, Braga 4710-057, Portugal}

\affiliation{$^{4}$Center for 2-Dimensional and Layered Materials, The Pennsylvania
State University, University Park, Pennsylvania 16802, USA}
\begin{abstract}
Adatom-decorated graphene offers a promising new path towards spintronics
in the ultrathin limit. We combine experiment and theory to investigate
the electronic properties of dilutely fluorinated bilayer graphene,
where the fluorine adatoms covalently bond to the top graphene layer.
We show that fluorine adatoms give rise to resonant impurity states
near the charge neutrality point of the bilayer, leading to strong
scattering of charge carriers and hopping conduction inside a field-induced
band gap. Remarkably, the application of an electric field across
the layers is shown to tune the resonant scattering amplitude from
fluorine adatoms by nearly twofold. The experimental observations
are well explained by a theoretical analysis combining Boltzmann transport
equations and fully quantum-mechanical methods. This paradigm can
be generalized to many bilayer graphene--adatom materials, and we
envision that the realization of electrically tunable resonance may
be a key advantage in graphene-based spintronic devices.
\end{abstract}


\maketitle
Chemical functionalization is a powerful tool to engineer graphene
for a broad range of application needs. The attachment of oxygen groups
makes graphene soluble and fluorescent, and facilitates the formation
of graphene nanocomposites \cite{01.Ruoff_06,02.Kikkawa_09,03.Chhowalla_10}.
Chemisorbed species increase the degree of $sp^{3}$ bonding, drastically
affecting graphene's electrical conductivity, mechanical strength,
and optical response, even in the very dilute limit \cite{05.Geim_09,16.Yazyev_10,06.Zhu_10,07.Hao_10,08.Snow_10,09.Geim_10,10.Geim_10b,11.Peeters_10,Fuhrer09,12.Eriksson_10,17.Zhu_11,13.Lau_11,14.Craciun_11,18.Zhu_12,19.Grigorieva_12,Fuhrer11,20.Kawakami_12,21.Ozyilmaz_13,15.Zhu_14}.
The deposition of light adatoms, such as hydrogen and fluorine, endows
graphene with intriguing magnetic and spintronic properties, including
localized magnetic moments and enhanced spin--orbit coupling, which
may enable the generation of large spin currents \cite{16.Yazyev_10,17.Zhu_11,18.Zhu_12,19.Grigorieva_12,20.Kawakami_12,21.Ozyilmaz_13,Ferreira_14,Ozyilmaz_Ferreira_14}.

Covalently bonded monovalent species, with orbitals close to the Dirac
point of graphene systems, effectively decouple their carbon host
from its neighbors, thereby simulating vacancies \cite{23.Lichtenstein}.
The latter are predicted to introduce power-law localized midgap states
at the Dirac point \cite{Pereira_07,Pereira_08,EVCastro10}, displaying
anomalous divergent behavior of the density of states \cite{Evers14,Ostrovsky}.
These so-called resonant scatterers have a profound impact on charge
carrier transport at all carrier densities, invalidating conventional
transport pictures based on the weak disorder hypothesis \cite{22.Guinea_07,24.ResScatt_mlg_blg,ZEM}.
In the high carrier density regime $|n|\gg n_{\textrm{rs}}$, where
$n_{\textrm{rs}}$ denotes the areal density of resonant scatterers,
the dc conductivity deviates from its typical behaviour $\sigma\propto|n|$,
acquiring a robust sublinear dependence, owing to a nonperturbative
renormalization of $s$-wave phase shifts: $\delta_{0}(n)\approx\pi/[2\ln(R\sqrt{\pi|n|})]$,
where $R\approx0.4$ nm is the scatterer range \cite{18.Zhu_12,22.Guinea_07,23.Lichtenstein,24.ResScatt_mlg_blg}.

Resonant scattering also plays an important role in metal spintronics
\cite{Fert 2011}, although such interactions are not easily tunable.
Engineering tunable resonant interaction of charge carriers with atomic
impurities in graphene \emph{in situ} would open new avenues, including
the harnessing of spin relaxation \cite{Fabian14} and the generation
of robust spin currents through the extrinsic spin Hall effect \cite{Ferreira_14,Ozyilmaz_Ferreira_14}.
From the viewpoint of device scaling and operations, it is particular
desirable to implement gate-controlled resonant interactions. A suitable
candidate---so far overlooked---is bilayer graphene (BLG). Adatoms
in BLG are likewise predicted to induce resonant scattering \cite{24.ResScatt_mlg_blg,27.Chetty_12,28.Dasgupta_13}.
Moreover, the presence of two layers allows the mirror symmetry to
be broken by an electric field perpendicular to the layers, opening
up a band gap up to 250~meV \cite{29.McCann_06,30.MacDonald_07,31.Wang_09,32.Heinz_09,34.Zhu_10b}.
The consequences of this extra degree of freedom for resonant scattering
remain largely unexplored.

In this Rapid Communication, we report experimental observations of
adatom-limited charge transport in BLG using fluorine adatoms as an
example. Dual gating allows us to control the carrier type, density,
and the perpendicular (bias) electric field independently. Both experiment
and theory demonstrate that fluorine adatoms behave as resonant scatterers.
Moreover, we show that the charge carrier scattering amplitude becomes
strongly electron-hole asymmetric and is tunable over a large range
$\pm20-30\%$ by controlling the carrier distribution between the
two layers using a bias electric field. Once more, theory and experiment
are found to be in excellent agreement. The demonstrated electric
tunability of the resonant cross section offers a convenient tool
to engineer desired charge--spin responses in graphene--adatom systems\emph{.}

\begin{figure}
\centering{}\includegraphics[width=1\columnwidth]{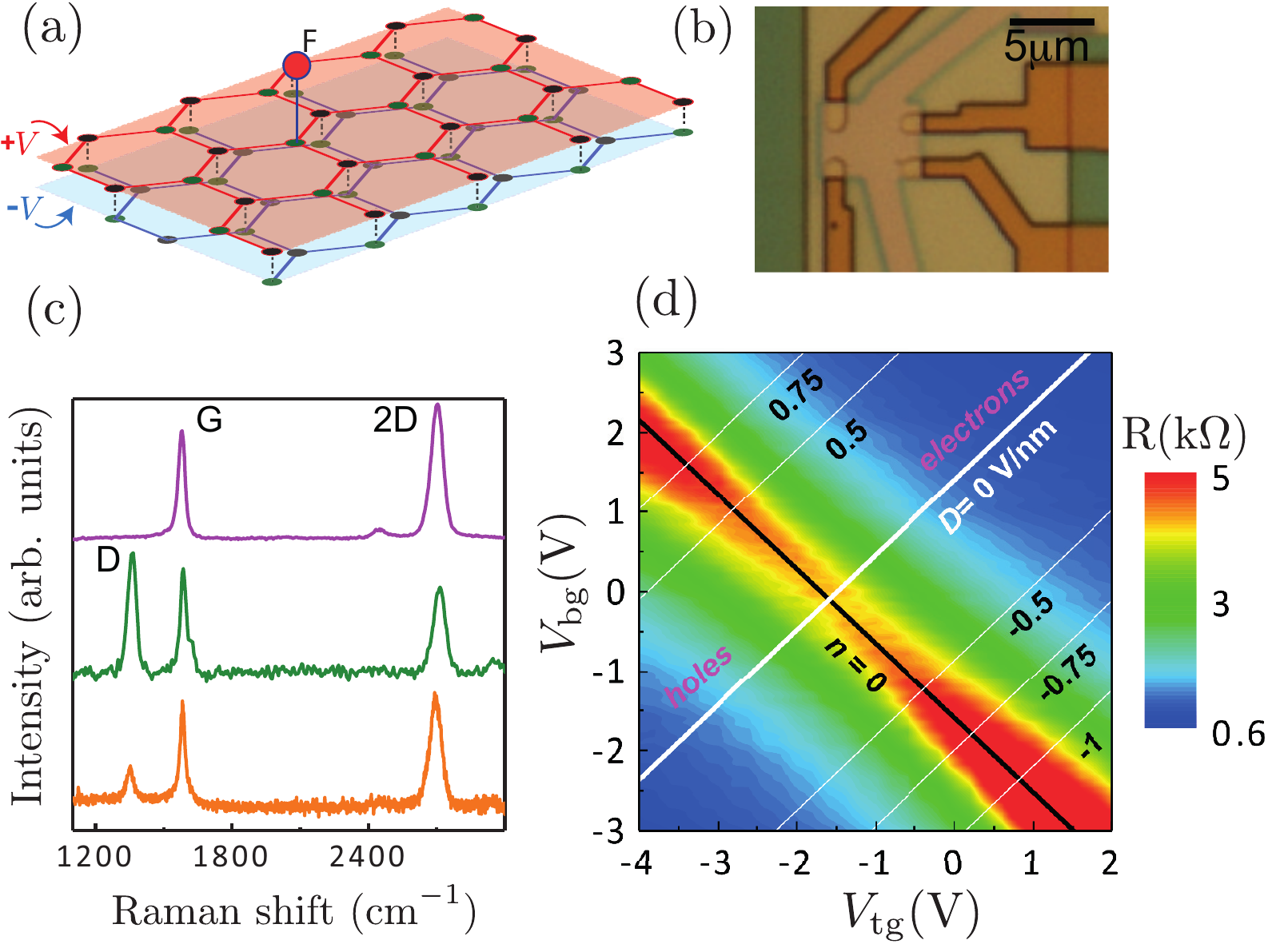}\protect\protect\protect\protect\protect\protect\protect\protect\caption{\label{fig:Schematics_and_DOS}Single-sided fluorinated BLG. (a) Schematic.
Solid (dashed) lines represent the intralayer (interlayer) hopping
terms. Fluorine adatoms bond to carbon hosts at random positions.
(b) Optical micrograph of sample W02 showing the Au bottom gate (wide
yellow stripe), four Au electrodes to the sample (dark gold) in van
der Pauw geometry, and the top gate connected to two Au electrodes
(yellow). (c) Raman spectra of pristine (plum), fluorinated (olive,
from sample W02) and defluorinated (orange) BLG. The laser excitation
wavelength is 488~nm. $I_{D}/I_{G}$=1.15 for W02 and 0.3 for the
defuorinated sample. (d) The color map of a four-terminal resistance
$R$ as a function of the top and bottom gate voltages. The black
line corresponds to $n$=0 and the white lines correspond to constant
bias $D$ fields. $T$=10 K. From W38. }
\end{figure}

Bilayer graphene flakes are exfoliated from bulk highly ordered pyrolytic
graphite (HOPG graphite) (ZYA grade, SPI supplies) onto prefabricated
HfO2/Au bottom gate stacks \cite{33.Zhu_Nano_13}, optically identified
and confirmed by Raman spectroscopy. The flakes are then fluorinated
in a CF$_{4}$ plasma asher where fluorine atoms covalently bond with
carbon atoms in the top layer only {[}see Fig.~\ref{fig:Schematics_and_DOS}(a){]}.
Bilayer graphene is substantially harder to fluorinate compared to
monolayer graphene \cite{17.Zhu_11}. Discussions of the plasma conditions
are given in the Supplemental Material (SM) \cite{SM}. After fluorination,
ver der Pauw or Hall bar devices are completed using standard $e$-beam
lithography and atomic layer deposition techniques \cite{34.Zhu_10b,33.Zhu_Nano_13}.
Results from three fluorinated samples (W02, W03, and W38) and one
defluorinated sample Df are reported here. An optical micrograph of
device W38 is shown in Fig.~\ref{fig:Schematics_and_DOS}(b). Fig.~\ref{fig:Schematics_and_DOS}(c)
plots the Raman spectra of a pristine bilayer, device W02, and a typical
trace from a defluorinated bilayer; see Supplemental Material and
Ref.~\cite{17.Zhu_11} for defluorination recipe. Following the atomic
defect density calibration obtained by Lucchese \emph{et al}. on monolayers~\cite{35.Jorio}
and taking into account the doubling of the G band intensity in bilayer,
we estimate a fluorine areal density $n_{\textrm{F}}^{\textrm{ram}}\approx3.3\times10^{12}$~cm$^{-2}$
for device W02, and the unintentional defect density (e.g., vacancy),
as seen in the defluorinated bilayer, is roughly 0.6$\times10^{12}$~cm$^{-2}$.
Parameters from all four samples are given in Table I of the Supplemental
Material. We use standard low-frequency, low excitation lock-in techniques
to carry out electrical transport measurements at temperatures ranging
from 1.6 K to 200 K. A false color map of a four-terminal resistance
$R$ in sample W38 as a function of the top and bottom gate voltages
$V_{\textrm{tg}}$ and $V_{\textrm{bg}}$ is shown in Fig.~\ref{fig:Schematics_and_DOS}~(d)
and displays characteristics similar to that of a pristine bilayer
\cite{34.Zhu_10b}. This map enables us to extract the dependence
of the sheet conductance $\sigma_{s}$ on the carrier density $n$
at fixed bias $D$~fields.

\emph{Resonant impurity scattering}.---We first establish the resonant
impurity nature of the fluorine adatoms. The rearrangement of electronic
spectral weight due to the fluorination can be estimated through a
calculation of the density of states (DOS) for a typical adatom concentration.
To model the BLG system we employ a nearest-neighbor tight-binding
Hamiltonian of $\pi$~electrons, supplemented with an on-site energy
term $\pm V$ in the top (bottom) layer describing the effect of the
$D$~field. Fluorine adatoms are modeled as vacancies in the top
layer \cite{23.Lichtenstein,24.ResScatt_mlg_blg}. The kernel polynomial
method \cite{KPM} is then used to accurately extract the DOS of a
large lattice with $2\times14\,142^{2}$ carbon atoms. Details of
the calculation are given in the Supplemental Material. 

\begin{figure}
\begin{centering}
\includegraphics[width=1\columnwidth]{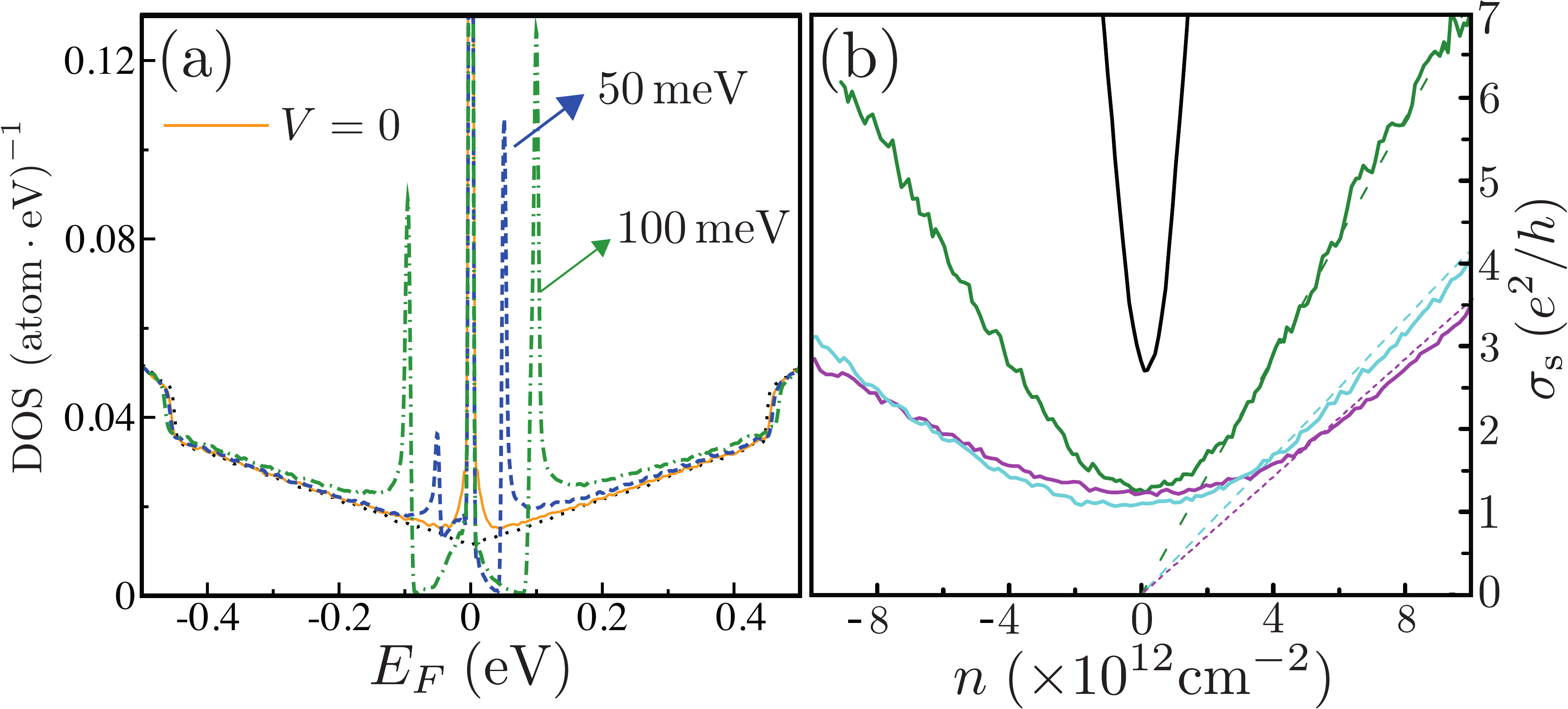} 
\par\end{centering}

\protect\protect\protect\protect\protect\protect\protect\protect\caption{\label{fig:Conduc_Low_Temp}Resonant impurity scattering in fluorinated
BLG. (a) Calculated DOS of macroscopic size ($\approx\mu$m$^{2}$)
BLG system as a function of the Fermi energy at selected bias potential
values for a F\,:\,C ratio of 1\,:\,2000. The DOS of pristine
unbiased BLG is also shown (black dotted line). (b) $\sigma_{\textrm{s}}(n)$
in samples (from top to bottom): Df (black), W38 (olive), W02 (cyan)
and W03 (plum) at $T=1.6$~K. Positive $n$ corresponds to electron
doping. Dashed lines of the same color are fits to Eq. (1). }
\end{figure}

The calculated DOS is shown in Fig.~\ref{fig:Conduc_Low_Temp}(a).
We note two prominent features. Firstly, in unbiased BLG, the DOS
displays a sharp peak centered at zero energy (midgap state), already
encountered in Refs.~\cite{24.ResScatt_mlg_blg,Yuan_10_BLG}. Secondly,
at nonzero bias field, new resonances pile up at the edges of the
pseudogap, which is consistent with results obtained for a single
vacancy \cite{EVCastro10}. Due to the broken layer symmetry, these
resonances have different weights at $E_{F}\approx\pm V$, anticipating
substantial electron-hole asymmetry (EHA). Fig.~\ref{fig:Conduc_Low_Temp}(b)
shows $\sigma_{\textrm{s}}(n)$ at $D=0$ for samples W38, W02, W03,
and Df at $T=$1.6~K. Sample Df has a field effect mobility of $\mu\approx$1600~cm$^{2}$/Vs,
which is comparable to that of a pristine bilayer in a similar geometry
\cite{34.Zhu_10b}. In contrast, fluorinated bilayers display much
lower mobility: 172, 100, and 86~cm$^{2}$/Vs for W38, W02, and W03,
respectively, which signals the dominance of adatom-induced scattering.
Ferreira \emph{et al}.~\cite{24.ResScatt_mlg_blg} showed that in
the high carrier density regime $n>n_{\textrm{F}}$ where quantum
corrections are not dominant, a semiclassical description applies,
and the resonant-scattering-limited sheet conductivity in unbiased
BLG is given by

\begin{equation}
\sigma_{\textrm{s}}(n)=2\times\frac{\pi e^{2}}{4h}\frac{|n|}{n_{\textrm{F}}}\,,\label{eq:sigma_res_scatt}
\end{equation}
where an extra factor of 2 accounts for the one-sided fluorination
of our samples. Since $\sigma_{\textrm{s}}$ is approximately insensitive
to the scatterer radius, there is a single fitting parameter, $n_{\textrm{F}}$.
Fits to Eq.~\ref{eq:sigma_res_scatt} are plotted as dashed lines
in Fig.~\ref{fig:Conduc_Low_Temp}(b) and describe data very well
at high density \cite{24.ResScatt_mlg_blg}. The extracted $n_{\textrm{F}}$
are in good agreement with values obtained from Raman spectra for
all samples (see Table I in Supplemental Material). Furthermore, in
all samples we find an $n_{F}$-independent $\sigma_{\textrm{s}}\approx e^{2}/h$
in the low carrier density regime $n<n_{\textrm{F}}$, also in agreement
with theory. This observation is consistent with the formation of
a narrow ``impurity band'' around the charge neutrality point (CNP)
\cite{24.ResScatt_mlg_blg,Yuan_10_BLG} {[}see thick (orange) line
in Fig.~\ref{fig:Conduc_Low_Temp}(a){]}.

\begin{figure}
\begin{centering}
\includegraphics[width=0.9\columnwidth]{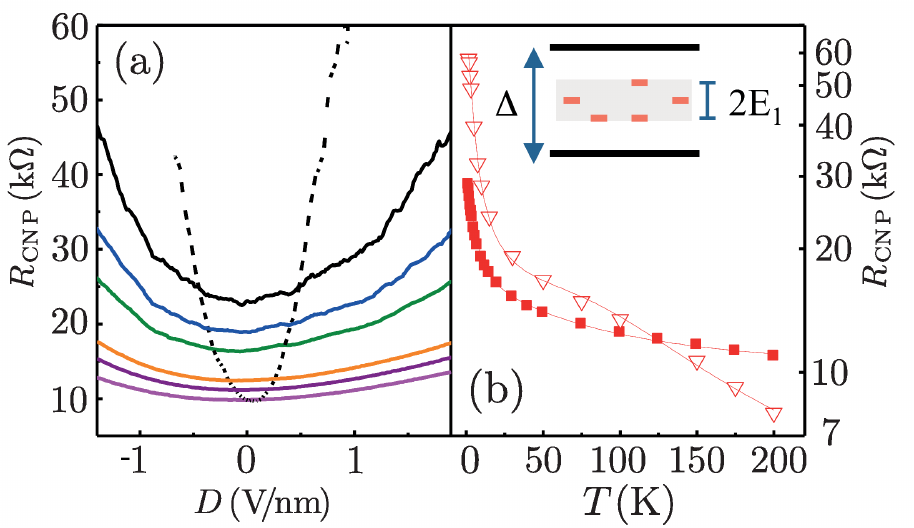} 
\par\end{centering}

\protect\protect\protect\protect\protect\protect\protect\protect\caption{\label{fig:COND_CNP}Charge carrier transport at the CNP. (a) $R_{s}$
vs $D$ at the CNP for sample Df (dashed line, $T$=1.6 K) and sample
W02 (solid lines). From top to bottom: $T$=1.6 K(black), 5 K (blue),
10 K (olive), 50 K(orange), 100 K(purple) and 200 K (plum). (b) $R_{S}(T)$
of sample Df (open triangles) and sample W02 (solid squares) at $D$=0.93
V/nm. The solid lines are fits to Eq.(1) of Ref.~\cite{34.Zhu_10b}.
$E_{1}$= 11 meV and 33.5 meV for samples W02 and Df, respectively.
Inset: A schematic diagram of the fluorine-induced impurity states
inside the field-induced band gap of a bilayer graphene.}
\end{figure}

More direct evidence of the impurity band is seen in the temperature-dependent
resistance of the CNP under a finite bias field. Figure~\ref{fig:COND_CNP}(a)
plots $R_{\textrm{CNP}}(D)$---the sheet resistance of the CNP in
sample W02---as a function of $D$ at selected temperatures $T=$1.6--200\,K,
together with $R_{\textrm{CNP}}(D)$ of sample Df at $T=$1.6\,K.
$R_{\textrm{CNP}}(D)$ of sample Df is similar to that of a pristine
bilayer \cite{34.Zhu_10b}. $R_{\textrm{CNP}}(D)$ of W02, on the
other hand, rises much more slowly, pointing to additional conduction
channels inside the band gap. To further understand the behavior of
$R_{\textrm{CNP}}$, we plot in Fig.~\ref{fig:COND_CNP}(b) its temperature
dependence in both samples at $D=$0.93 V/nm. The most remarkable
difference between the two lies at high temperatures $T>50$~K. In
this regime, sample Df exhibits activated transport $R_{1}\propto\exp(E_{1}/k_{B}T)$,
with $E_{1}$ increasing with $D$ from 22.5\,meV at $D=0$ (data
not shown) to 33.5\,meV at $D=0.93$\,V/nm. These values are similar
to pristine bilayers, where $\Delta=2E_{1}$ approximates the bias-induced
band gap in the large $D$ limit \cite{34.Zhu_10b}. In contrast,
similar analyses done on $R_{\textrm{CNP}}(T)$ of the three fluorinated
bilayers yield roughly $D$-independent $E_{1}$, which is approximately
11-13 meV in all samples. We attribute this behavior to nearest-neighbor
hopping among the fluorine-induced impurity states, as illustrated
in the inset to Fig.~\ref{fig:COND_CNP}(b), where $E_{1}$ is the
half-width of the impurity band. This mechanism effectively shunts
the band edge activation $\exp(\Delta/2k_{B}T)$ to result in a $D$-independent
$R_{\textrm{CNP}}(T)$. Counting two impurity states per adatom \cite{24.ResScatt_mlg_blg},
we independently estimate the bandwidth to be $\approx$10~meV at
$n_{\textrm{F}}\approx4\times10^{12}$~cm$^{-2}$, which is in excellent
agreement with the $E_{1}$ values extracted here.

\begin{figure}
\centering{}\includegraphics[width=0.9\columnwidth]{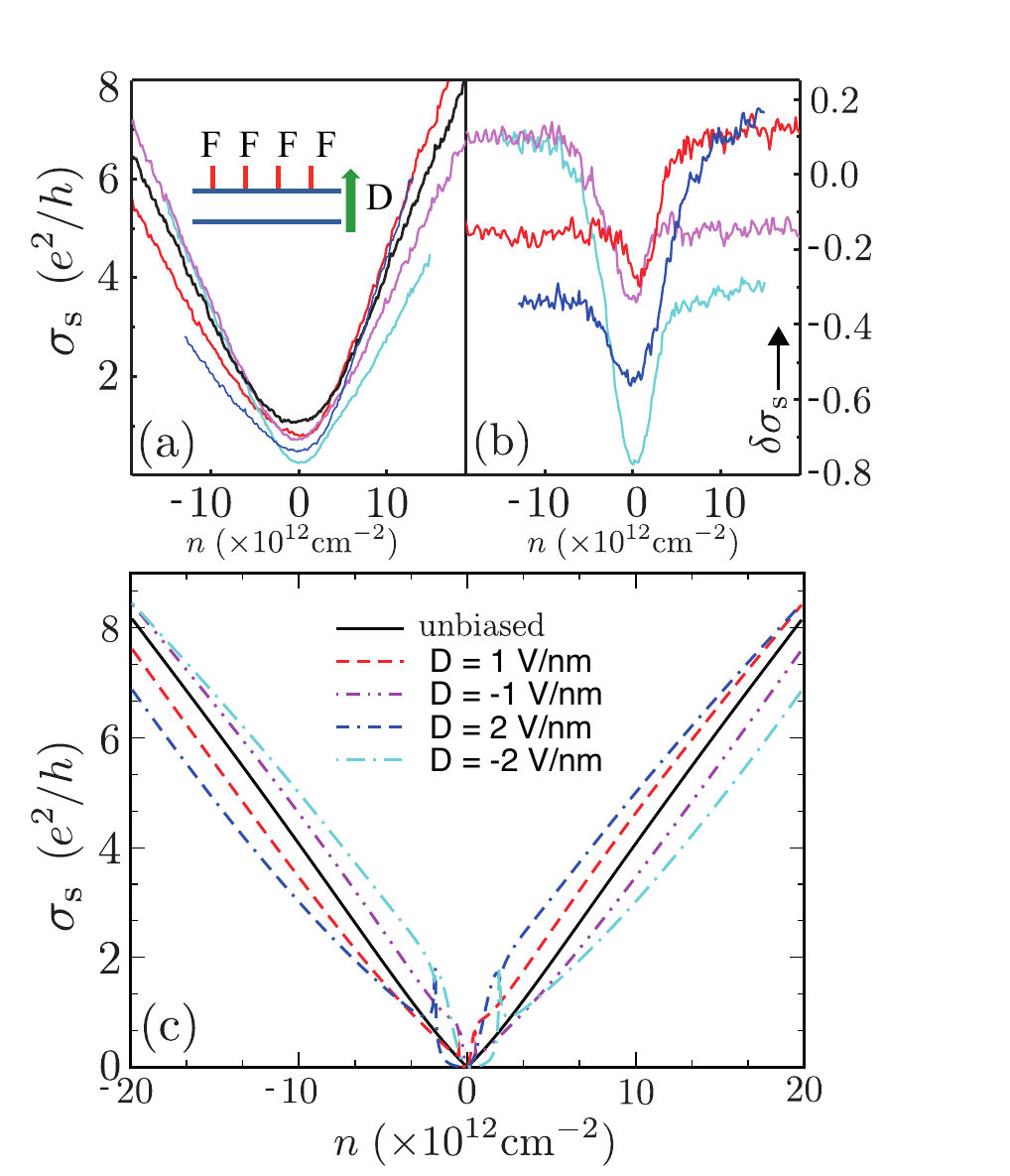}\protect\protect\protect\protect\protect\protect\protect\protect\caption{\label{fig:Cond_ZeroTemp_Theory}Tuning of charge carrier transport
in $D$-field. (a) and (b) $\sigma_{\textrm{s}}(n)$ and normalized
conductance change $\delta\sigma_{s}(D)$ for W02 at $T=1.6$ K and
$D=$-2~V/nm (cyan), -1~V/nm (plum), 0~V/nm (black), 1~V/nm (red),
and 2~V/nm (blue). Positive $D$ points towards the fluorine adatoms
as illustrated in the inset of(a). (c) Calculated dc-conductivity
$\sigma_{\textrm{s}}(n)$ for several inter-layer bias potentials
corresponding to the $D-$field values in (a). Calculation only applies
to $n>n_{F}$. Other parameters in main text. }
\end{figure}

\emph{Electric-field tuning of resonant scattering}.---Next, we explore
the role of a bias $D$~field in the carrier density regime $n>n_{\textrm{F}}$.
Figure~\ref{fig:Cond_ZeroTemp_Theory}(a) plots $\sigma_{\textrm{s}}(n)$
of sample W02 at selected $D$'s from -2 to 2~V/nm, where $D>0$
indicates field pointing towards the fluorine adatoms, as the inset
of (a) shows. Pronounced EHA is observed. For electrons, $\sigma_{s}$
increases (decreases) when \textbf{$D>0$ ($D<0$)} and the opposite
trend is observed for holes. This $D$-field tuning is further illustrated
in Fig.~\ref{fig:Cond_ZeroTemp_Theory}(b), where we plot the normalized
conductance change $\delta\sigma_{\textrm{s}}(D)\equiv\Delta\sigma_{\textrm{s}}(D)/\sigma_{\textrm{s}}(0)=[\sigma_{\textrm{s}}(D)-\sigma_{\textrm{s}}(0)]/\sigma_{\textrm{s}}(0)$
of the data in Figure~\ref{fig:Cond_ZeroTemp_Theory}~(a). Strikingly,
in the high-density regime, $\delta\sigma_{\textrm{s}}(D)$ is roughly
$n$-independent and ranges between -0.35 and 0.2 for $|D|<2$\,V/nm,
i.e., the change of $\sigma_{\textrm{s}}$ is nearly twofold. Furthermore,
$\delta\sigma_{\textrm{s}}(D)$ appears to be symmetric under $(n,D)\rightarrow(-n,-D)$.
Similar $D$-field tuning of $\sigma_{\textrm{s}}(n)$ is observed
on all three fluorinated samples, with the magnitude of $\delta\sigma_{\textrm{s}}(D)$
varying within a factor of 2 among them. Further examination of the
EHA rules out an extrinsic mechanism due to nonuniform density distribution
created by electrodes that screen a portion of the top gate. The detailed
discussions are given in the Supplemental Material. Intuitively, the
experimental observations suggest an increased weight of hole (electron)
wave functions in the top layer under a positive (negative) D field,
and thus increased resonant impurity scattering.

To gain further insight into the strong $D$-field tuning of charge
carrier transport, we solve analytically the Boltzmann transport equations
for a BLG system with random short-range impurities constrained to
one layer. We posit our investigations in the four-band continuum
BLG--adatom Hamiltonian 
\begin{equation}
\hat{H}=\left(\begin{array}{cc}
V+v_{F}\boldsymbol{\sigma}\cdot\hat{\mathbf{p}} & t_{\perp}\sigma_{-}\\
t_{\perp}\sigma_{+} & -V+v_{F}\boldsymbol{\sigma}\cdot\hat{\mathbf{p}}
\end{array}\right)+\sum_{i}\left(\begin{array}{cc}
v_{0}^{i}(\mathbf{r}) & 0\\
0 & 0
\end{array}\right)\,,\label{eq:H_total}
\end{equation}
where $\hat{\mathbf{p}}$ is the two-dimensional (2D) kinematic momentum
operator around Dirac point $K$, $v_{F}\approx10^{6}$\,m/s is the
Fermi velocity, $\boldsymbol{\sigma}$ are Pauli matrices, with $\sigma_{z}=\pm1$
describing states residing on the $A$($B$) sublattice, and $\sigma_{\pm}=\sigma_{x}\pm i\sigma_{y}$.
The diagonal blocks in the first term on the right-hand side describe
monolayers at a constant electrostatic energy, $\pm V$ for top layer
and bottom layer, respectively, whereas off-diagonal blocks contain
the hopping processes $A_{2}\rightleftharpoons B_{1}$ connecting
the two layers ($t_{\perp}\approx0.45$~eV) \cite{EVCastro_Review}.
The bias potential $V$ is determined from the experimental parameters
according to $V=-eDd/2\kappa$, where $-e<0$ is the electron's charge,
$d\approx0.35$ nm is the layer separation, and $\kappa=4$ is the
phenomenological BLG effective permittivity \cite{30.MacDonald_07,screening}.
Diagonalization of the first term yields $2+2$ bands separated by
a gap: $\Delta=2|V|t_{\perp}/\sqrt{t_{\perp}^{2}+4V^{2}}\approx2|V|$.
Their dispersion relation reads $\epsilon^{\pm\pm}(\boldsymbol{\pi})=\pm\sqrt{t_{\perp}^{2}/2+\boldsymbol{\pi}^{2}+V^{2}\pm\lambda^{2}(\boldsymbol{\pi})}$,
where $\boldsymbol{\pi}\equiv v_{F}\mathbf{p}$ and $\lambda^{2}(\boldsymbol{\pi})=\sqrt{t_{\perp}^{4}/4+\boldsymbol{\pi}^{2}\left(t_{\perp}^{2}+4V^{2}\right)}$.
The second term is the scattering potential due to fluorine adatoms
located in the top layer at random positions $\{\mathbf{r}_{i}\}$
with $v_{0}^{i}(\mathbf{r})=v_{0}\delta(\mathbf{r}-\mathbf{r}_{i})$.
In the scattering problem, outgoing and incoming wavefunctions are
related by the $T$~matrix \cite{Lippmann-Schwinger}. For delta-peak
potentials, outgoing waves from a single impurity at the origin acquire
the simple form $\psi_{\textrm{scatt}}(\mathbf{r})=\hat{G}_{0}(\mathbf{r},E)\hat{T}_{\textrm{ad}}(E)\phi_{\mathbf{k}}(\mathbf{r})$,
where $\phi_{\mathbf{k}}(\mathbf{r})$ denotes free incoming wave
solutions with momentum $\mathbf{k}$, i.e., $\hat{H}_{0}|\phi_{\mathbf{k}}\rangle=E|\phi_{\mathbf{k}}\rangle$,
and $\hat{G}_{0}(\mathbf{r},E)$ is the propagator of pristine biased
BLG, $\hat{G}_{0}(\mathbf{r},E)=\langle\mathbf{r}|(E-\hat{H}_{0}+i\eta)|\vec{0}\rangle$,
with $\eta$ a real infinitesimal. In the (resonant) scattering limit
of interest $v_{0}\rightarrow\infty$, we easily find 
\begin{equation}
\hat{T}_{\textrm{ad}}(E)=-\sum_{\chi=A_{2},B_{2}}[g_{\chi}(E)]^{-1}|\chi\rangle\langle\chi|\,,\label{eq:T-matrix}
\end{equation}
where $g_{\chi}(E)\equiv\langle\chi|\hat{G}(\vec{0},E)|\chi\rangle$.
We evaluated $g_{\chi}(E)$ in the entire parameter space (refer to
Supplemental Material); in the intermediate regime $\sqrt{t_{\perp}^{2}+V^{2}}>|E|>|V|$
(typically $\sim$0.05--0.5 eV), 
\begin{align}
g_{A_{2}}(E) & =(V-E)\left[\bar{\Theta}_{\Lambda}(E)+(E+V)^{2}\Theta_{\textrm{reg}}(E)\right]\,,\label{eq:g_A}\\
g_{B_{2}}(E) & =(t_{\text{\ensuremath{\perp}}}^{2}+V^{2}-E^{2})(E+V)\Theta_{\textrm{reg}}(E)\nonumber \\
 & \quad+(V-E)\bar{\Theta}_{\Lambda}(E)\,,\label{eq:g_B}
\end{align}
where 
\begin{align}
\left\{ \begin{array}{c}
\Theta_{\textrm{reg}}(E)\\
\bar{\Theta}_{\Lambda}(E)
\end{array}\right\}  & =\frac{1}{4\pi v_{F}^{2}}\frac{1}{A_{+}+A_{-}}\left\{ \begin{array}{c}
\ln\left(\frac{A_{+}}{A_{-}}\right)-i\pi\\
\Psi\left(\Lambda\right)+i\pi A_{+}
\end{array}\right\} \,.\label{eq:Theta_Reg}
\end{align}
In the above, $A_{\pm}=\sqrt{4E^{2}V^{2}+t_{\perp}^{2}(E^{2}-V^{2})}\pm(E^{2}+V^{2})$,
$\Psi(x)=\sum_{p=\pm1}A_{p}\ln(-p+x^{2}/|A_{p}|)$, and $\Lambda=\hbar v_{F}/R$
is the high-energy cutoff defining an effective potential range $R$
\cite{24.ResScatt_mlg_blg}. Crucially, for non-zero bias field $|V|>0$,
the $T$~matrix is sensitive to carrier polarity $\pm|E|$ {[}see
Eqs.~(\ref{eq:g_A})--(\ref{eq:g_B}){]}. This feature results from
the top-bottom layer asymmetry induced by the adatoms and, as argued
below, is the origin of the strong EHA observed.

The effect of dilute random adatoms in dc-transport is encoded in
the inverse transport relaxation time 
\begin{align}
\frac{1}{\tau(\mathbf{k})} & =\frac{2\pi n_{\textrm{F}}}{\hbar}\int\frac{d^{2}\mathbf{k}^{\prime}}{(2\pi)^{2}}|\langle\phi_{\mathbf{k}}|\hat{T}_{\textrm{ad}}(E)|\phi_{\mathbf{k}^{\prime}}\rangle|^{2}\delta_{\mathbf{k}\mathbf{k}^{\prime}}(E)\,,\label{eq:Sigma_Transp}
\end{align}
where $\delta_{\mathbf{k}\mathbf{k}^{\prime}}(E)\equiv(1-\cos\theta_{\mathbf{k}\mathbf{k}^{\prime}})\delta(\epsilon_{\mathbf{k}}-\epsilon_{\mathbf{k}^{\prime}})$,
$\theta_{\mathbf{k}\mathbf{k}^{\prime}}$ is the scattering angle,
and $\epsilon_{\mathbf{k}}\equiv\epsilon^{\pm-}(v_{F}\hbar\mathbf{k})$
for $E=\pm|E|$. The zero-temperature semiclassical dc-conductivity
follows from $\sigma_{\textrm{s}}=(e^{2}/h)k_{F}v(k_{F})\tau(k_{F})$,
where $v(k)=|\nabla_{k}\epsilon_{k}|/\hbar$ is the band velocity
and $k_{F}$ is the Fermi wave vector~\cite{technical_detail}. In
order to unbiasedly extract the model parameters $n_{\textrm{F}}$
and $R$, we fit the 4-band expression to the $D=0$ conductivity
data in the electron sector $n>0$ {[}see black curve in Fig.~\ref{fig:Cond_ZeroTemp_Theory}~(a){]}.
We obtain $n_{\textrm{F}}\approx1.6n_{\textrm{F}}^{\textrm{ram}}$
and $R\approx0.5$~nm, where $n_{\textrm{F}}^{\textrm{ram}}$ is
obtained from the Raman spectra of this device (see above). These
parameters are then employed to evaluate $\sigma_{\textrm{s}}(n)$
at all nonzero $D$-fields in Fig.~\ref{fig:Cond_ZeroTemp_Theory}~(c).
The theory is seen to reproduce very well both the degree of EHA and
the overall $D$-dependence observed in panel (a). This agreement
is particularly remarkable as the strong scattering regime is characterized
by a strong dependence of $\tau^{-1}(\mathbf{k})$ in the $D$-field
and model parameters. This robust agreement is strong evidence that
the $D$-field tuning of the resonant adatom scattering amplitudes
(\ref{eq:T-matrix}) is responsible for the observed conductance modulation.
Because of the resonant nature of the adatoms, a twofold change of
the scattering cross section can lead to very large tuning of other
resonant properties, such as the spin Hall angle \cite{Ferreira_14},
giving BLG-based systems a distinct advantage in graphene-based spintronics.

In summary, through a joint experiment--theory effort, we have demonstrated
that a perpendicular electric field achieves substantial tuning of
the amplitude of resonant impurity scattering in one-sided fluorinated
bilayer graphene. Our findings set the stage for exploring all-electric
control of resonant scatterings that underlie novel spintronics effects
in graphenic materials.

\emph{Acknowledgements}. We thank X. Hong for helpful discussions.
A.S., J.L., and J.Z. are supported by ONR under Grant No.~N00014-11-1-0730
and by NSF CAREER Grant No.~DMR-0748604. A.F. and N.M.R.P. acknowledge
EC under Graphene Flagship (Contract No.~CNECT-ICT-604391). A.F.
gratefully acknowledges the financial support of the Royal Society
(U.K.) through a Royal Society University Research Fellowship. We
acknowledge use of facilities at the PSU site of NSF NNIN.

\begin{center}
\textbf{\large{}{}{}{}{}{{Supplemental Material for ``}}Electrically
tunable resonant scattering in fluorinated bilayer graphene{}{}''}{\large{}{}{}{}{}{{{}
\setcounter{equation}{0}}}} {\large{}{}{}{}{}{{\setcounter{figure}{0}}}} 
\par\end{center}

We give details on the fluorination and characterization of the biased
bilayer graphene devices, and provide comprehensive derivations of
the main results discussed in the Rapid Communication. Additional
material includes explicit formulae for scattering properties such
as cross sections, and fully quantum-mechanical tight-binding calculations
in very large disordered systems. In particular, we show that the
dependence of the Kubo dc-conductivity with the bias field is in good
qualitative agreement with the predictions from the four-band semiclassical
calculation.

\tableofcontents{}

\section{Fabrication and Characterization of Fluorinated Bilayer Graphene
Devices\label{sec:Fabrication-and-Characterization}}

\subsection{Fluorination of Bilayer Graphene}

All samples were fluorinated in a plasma asher (Metroline M4L) using
CF$_{4}$. Defluorination was done by annealing the sample in Ar/H$_{2}$
flow at 360 $^{\circ}$C for 12 hours. Compared to monolayer graphene,
bilayer graphene is substantially harder to fluorinate, presumably
due to a flatter topography. Three parameters, namely, the power p,
the gas pressure P and the duration time T, control the fluorine coverage
as well as the unintentional vacancy density. Empirical trials show
that the ratio of fluorine density vs vacancy density is maximized
to be roughly 4:1 under conditions of p=100 W and P=200mTorr (100
sccm of CF$_{4}$ gas flow). At this setting, a T=5 min run produces
approximately $n_{\textrm{F}}$=$1\times10^{12}\text{cm}^{-2}$.

\subsection{Color Map of 4-Terminal Resistance}

We attribute the resistance peak {[}black line in Fig.~1(d), main
text{]} to the charge neutrality point (CNP). Tracking its evolution
with $V_{\textrm{tg}}$, $V_{\textrm{bg}}$ and the location of a
minimum along the black line allows us to determine the gating efficiencies
of the top and bottom gates $\alpha_{\textrm{tg}}$ and $\alpha_{\textrm{bg}}$,
respectively, and also the unintentional doping of the two gates,
$V_{\textrm{tg}}^{0}$ and $V_{\textrm{bg}}^{0}$, respectively. The
total carrier density is $n=\alpha_{\textrm{tg}}(V_{\textrm{tg}}-V_{\textrm{tg}}^{0})+\alpha_{\textrm{bg}}(V_{\textrm{bg}}-V_{\textrm{bg}}^{0})$,
and the electric displacement field is $D=(D_{\textrm{tg}}+D_{\textrm{bg}})/2$,
where $D_{\textrm{bg}}=(\alpha_{\textrm{bg}}e)/\varepsilon_{0}(V_{\textrm{bg}}-V_{\textrm{bg}}^{0})$
and $D_{\textrm{tg}}=-(\alpha_{\textrm{tg}}e)/\varepsilon_{0}(V_{\textrm{tg}}-V_{\textrm{tg}}^{0})$.
Table~\ref{tab:Device-characteristics} lists the parameters of all
devices used in this study.

\begin{widetext}

\begin{table}[H]
\begin{centering}
\begin{tabular}{|ccccc|}
\hline 
 & \quad{}\quad{}W38\quad{}\quad{}  & \quad{}\quad{}W02\quad{}\quad{}  & \quad{}\quad{}W03\quad{}\quad{}  & \quad{}\quad{}Df\quad{}\quad{}\quad{}\tabularnewline
\hline 
$\alpha_{\textrm{tg}}$($\times10^{12}$~cm$^{-2}$V$^{-1}$)  & 2.3  & 3.15  & 3.15  & 2.3\tabularnewline
$\alpha_{\textrm{bg}}$($\times10^{12}$~cm$^{-2}$V$^{-1}$)  & 2.44  & 3  & 3  & 0.065\tabularnewline
\hline 
$V_{\textrm{tg0}}$~(V)  & -1.7  & -3.7  & -6  & -1.3\tabularnewline
$V_{\textrm{bg0}}$~(V)  & -0.2  & 0.6  & -3.1  & -7.7\tabularnewline
\hline 
$I_{D}/I_{G}$  & 1.15  & 1.15  & 1.3  & 0.3\tabularnewline
$I_{D}/I_{G}${*}{*}  & 0.9  & 1.25  & 1.4  & N.A.\tabularnewline
\hline 
$n_{\textrm{F}}$($\times10^{12}$~cm$^{-2}$)  & 2.2  & 3.8  & 4.4  & 0.6\tabularnewline
\hline 
\end{tabular}
\par\end{centering}

\protect\protect\protect\caption{\label{tab:Device-characteristics}Device characteristics for 3 dilute
fluorinated bilayer graphene samples (W38, W03, and W02) and one defluorinated
bilayer graphene sample (Df). Parameters are described in the text.
ID/IG{*}{*} indicates the expected ID/IG from resonant scattering
fits. }
\end{table}

\end{widetext}

\subsection{Electron--Hole Assymetry}

\begin{figure}
\begin{centering}
\includegraphics[width=1\columnwidth]{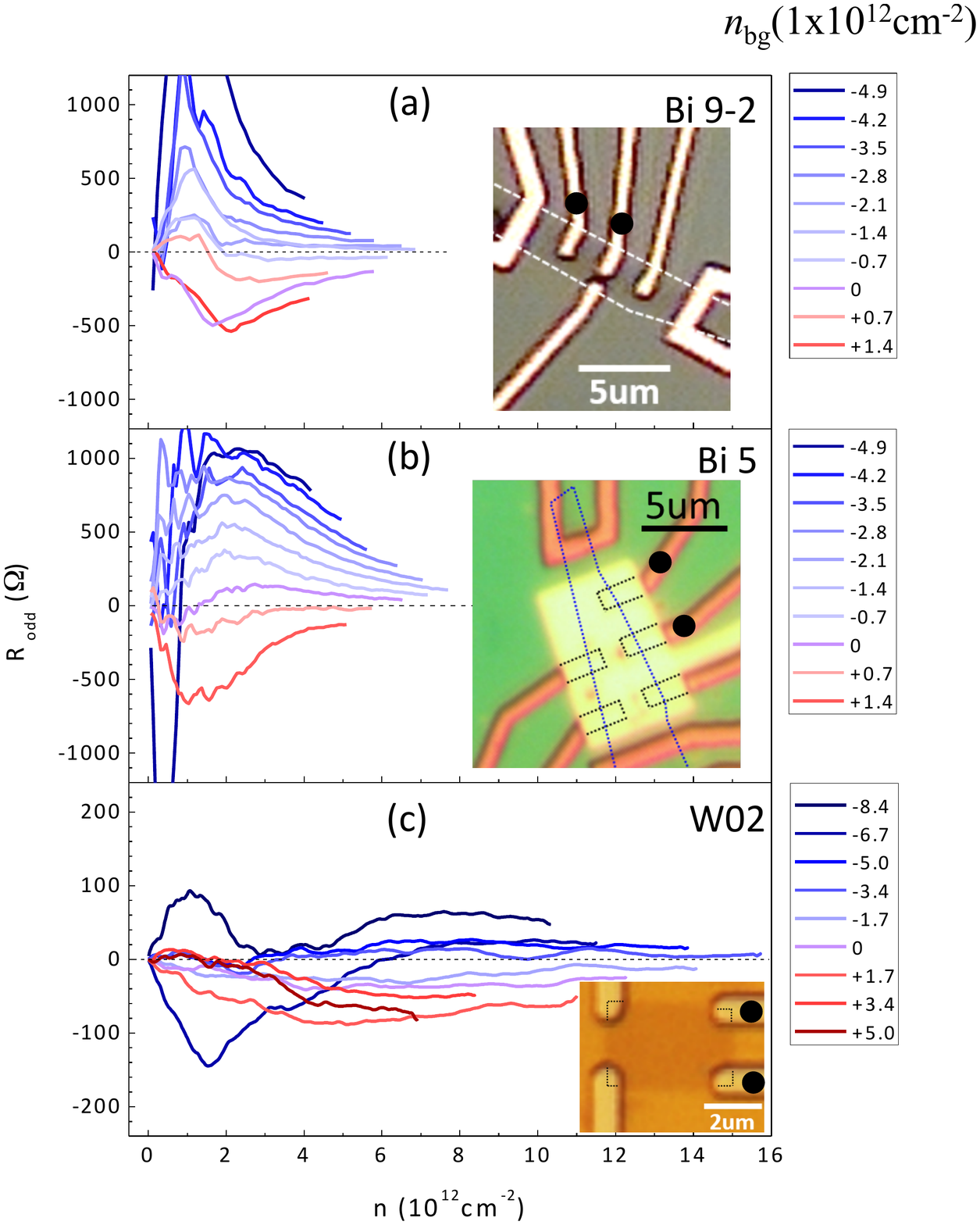} 
\par\end{centering}

\protect\protect\caption{\label{fig:R_odd}A close look into the electron--hole asymmetry.
$R_{\textrm{odd}}$ vs electron density $n$ at selected $n_{\textrm{bg}}$
for pristine bilayers Bi 9-2, Bi5 and fluorinated bilayer W02. (Bi9-2
and Bi 5 were reported in Ref.~\cite{Zou10}). The magnification
of the y-axis is adjusted to account for the geometry difference between
the three measurements, ensuring a fair comparison. Positive $n_{\textrm{bg}}$
corresponds to electrons. Insets are optical micrographs of the devices.
The graphene flake is outlined in each image. Black dots indicate
voltage probes used. In Bi9-2 and Bi 5, the voltage probes are fairly
invasive with pronounced junction effect. They are minimally invasive
in W02. }
\end{figure}

The density difference between the dual-gated portion of a bilayer
and the area underneath the metal electrodes, whose density is only
controlled by the backgate, can lead to bipolar junctions that are
more resistive than unipolar junctions. Similar phenomena were studied
in monolayers~\cite{Huard2008}. In our dual-gated bilayer, increasing
$V_{\textrm{bg}}$ in the positive direction creates a n-doped region
underneath the contact, which favors the conduction of electrons.
This extrinsic electron--hole asymmetry carries the same sign as the
$D$-field controlled resonant impurity scattering discussed in the
main text since a positive $V_{\textrm{bg}}$ results in a positive
$D$-field so care must be taken to differentiate between the two.

The junction induced electron--hole asymmetry becomes less important
as the contact become less intrusive. Compared to conventional bar-like
devices (pristine Bi 9-2 and Bi 5 shown in Fig.~\ref{fig:R_odd}),
fluorinated bilayer W02 shown in the main text {[}Fig.~1(c){]} is
a van der Pauw geometry with minimally intrusive voltage probes on
the current path. Naturally, we expect a small impact from the junction
effect on W02. In Fig.~\ref{fig:R_odd}, we plot $R_{\textrm{odd}}$
vs electron density $n$ of the dual-gated area at fixed $n_{\textrm{bg}}$,
using 
\[
R_{\textrm{odd}}(n_{\textrm{bg}})=\frac{1}{2}\left[R(n,n_{\textrm{bg}})-R(-n,n_{\textrm{bg}})\right]\,,
\]
as a measure of the electron--hole asymmetry \cite{Huard2008}. Fig.~\ref{fig:R_odd}(a)-(c)
compare results from two pristine bilayer samples and the fluorinated
W02 (in Fig.~4, main text). Bi 9-2 and Bi 5 use 300nm SiO$_{2}$
backgates so the carrier density and $D$ field range are smaller
than that of W02.

On the pristine devices, $R_{\textrm{odd}}$ increases with the magnitude
of $n_{\textrm{bg}}$ and reverses sign in the vicinity of $n_{\textrm{bg}}=0$,
as expected from the junction effect. Furthermore, $R_{\textrm{odd}}$
decreases with increasing $n$ and tends toward zero at high $n$.
In comparison, $R_{\textrm{odd}}$ in W02, which is non-zero because
it does exhibit electron--hole asymmetry, does not correlate with
$n_{\textrm{bg}}$ and shows no dependence on $n$. This rules out
the junction effect as the origin of the reported electron--hole asymmetry
tuning.

\section{Solution of the Scattering Problem\label{sec:Scatt_Amp}}

\subsection{Propagator of Biased Bilayer Graphene\label{sec:Propagator}}

In this section we provide the derivation of the propagator of low-energy
quasi-particles in Bernal-stacked bilayer graphene in the presence
of a perpendicular electric field. For our purposes it is sufficient
to consider the following minimal Hamiltonian describing $\pi$-electrons
in the $K$ valley \cite{BLG} 
\begin{equation}
\hat{H}_{K}={\color{red}\begin{array}{c}
{\color{blue}{\color{cyan}{\color{red}\stackrel{{\color{blue}B_{2}}}{}}}}\\
{\color{blue}{\color{cyan}{\color{red}\stackrel{{\color{blue}A_{1}}}{}}}}\\
{\color{blue}{\color{cyan}{\color{red}{\color{red}\stackrel{{\color{blue}B_{1}}}{}}}}}\\
{\color{blue}{\color{cyan}{\color{red}\stackrel{{\color{blue}A_{2}}}{}}}}
\end{array}}\underset{{\color{red}{\color{blue}B_{2}\quad\quad A_{1}\quad\quad B_{1}\quad\quad A_{2}}}}{\left(\begin{array}{cccc}
V & 0 & 0 & \hat{\pi}\\
0 & -V & \hat{\pi}^{\dagger} & 0\\
0 & \hat{\pi} & -V & -t_{\perp}\\
\hat{\pi}^{\dagger} & 0 & -t_{\perp} & V
\end{array}\right)}\,,\label{eq:Hamiltonian_Bernal_BLG}
\end{equation}
where $\hat{\pi}\equiv v_{F}(\hat{p}_{x}+i\hat{p}_{y})$ {[}here $\hat{\mathbf{p}}=-(i/\hbar)\boldsymbol{\nabla}$
is the 2D momentum operator and $v_{F}\approx10^{6}$~ms$^{-1}$
is the Fermi velocity{]}, $t_{\perp}$ denotes the inter-layer hopping,
and $V$ is half the energy difference between the layers as induced
by a perpendicular electric field \cite{fine_features}. Hereafter
we set $\hbar\equiv1$. The dispersion relations of the four bands
associated with $\hat{H}_{K}$ read as 
\begin{equation}
\epsilon^{\pm\pm}(\boldsymbol{\pi})=\pm\sqrt{\frac{t_{\perp}^{2}}{2}+\boldsymbol{\pi}^{2}+V^{2}\pm\sqrt{\frac{t_{\perp}^{4}}{4}+\boldsymbol{\pi}^{2}\left(t_{\perp}^{2}+4V^{2}\right)}}\,,\label{eq:dispersion_relation}
\end{equation}
where $\boldsymbol{\pi}=v_{F}\mathbf{p}\equiv v_{F}p(\cos\theta_{\mathbf{p}},\sin\theta_{\mathbf{p}})^{\textrm{t}}$.
The presence of a bias potential opens a gap $\Delta=2|Vt_{\perp}|/\sqrt{4V^{2}+t_{\perp}^{2}}$.
Usually $|V|$ is small, and thus $\Delta\simeq2|V|$ is a good approximation.
Depending on the position of the Fermi level, three spectral regions
must be considered: 
\begin{itemize}
\item Region I: $|V|>|E|>\Delta/2$, the ``Mexican hat'' region. Here,
the Fermi level crosses the band $\epsilon^{\pm-}$ twice. 
\item Region II: $E_{H}>|E|>\Delta/2$, the intermediate density region
where the Fermi level crosses $\epsilon^{\pm-}$ at a single point. 
\item Region III: $\Lambda>|E|>E_{H}$ , the high-energy region where the
Fermi level crosses $\epsilon^{\pm\pm}$ and $\epsilon^{\pm-}$ at
a single point. 
\end{itemize}
In the above, $\pm E_{H}=\pm\sqrt{t_{\perp}^{2}+V^{2}}$ denotes the
maximum (minimum) of the band $\epsilon^{\pm+}$ and $\Lambda$ a
cutoff of the low-energy theory (see below). In what follows we compute
the propagator in coordinate space in regions I--III.

The resolvent operator $\hat{G}(z)\equiv(z-\hat{H}_{K})^{-1}$ has
a simple representation in momentum space:
\begin{align}
\mathcal{G}(z,\mathbf{\boldsymbol{\pi}})= & \left[z-\frac{t_{\perp}}{2}\left(\sigma_{z}\otimes\sigma_{x}-\sigma_{0}\otimes\sigma_{x}\right)\right.\nonumber \\
 & \left.-|\mathbf{\boldsymbol{\pi}}|\sigma_{x}\otimes(\mathbf{n}^{\prime}\cdot\boldsymbol{\mathbf{\sigma}})-V\sigma_{z}\otimes\sigma_{z}\right]^{-1}\,,\label{eq:equation_for_G}
\end{align}
 where $\mathbf{n}^{\prime}=(\cos\theta_{\mathbf{p}},-\sin\theta_{\mathbf{p}})^{\textrm{t}}$,
$\sigma_{\alpha}$ are Pauli matrices ($\alpha=x,y,z)$, and $\sigma_{0}$
is the $2\times2$ identity. The matrix inversion in (\ref{eq:equation_for_G})
is straightforward. We obtain 
\begin{eqnarray}
\mathcal{G}(z,\mathbf{\boldsymbol{\pi}}) & = & \frac{1}{D(z,\mathbf{\boldsymbol{\pi}})}\:\mathcal{D}^{(4)}(z,\mathbf{\boldsymbol{\pi}})\,,\label{eq:propagator-G}
\end{eqnarray}
where
\begin{align}
D(z,\mathbf{\boldsymbol{\pi}}) & \equiv-\prod_{\lambda,\lambda^{\prime}=\pm1}\left[z-\epsilon^{\lambda\lambda^{\prime}}(\mathbf{\boldsymbol{\pi}})\right]\nonumber \\
 & =-\mathbf{\boldsymbol{\pi}}{}^{4}+2\mathbf{\boldsymbol{\pi}}^{2}(z^{2}+V^{2})\nonumber \\
 & \quad+(z^{2}-V^{2})(t_{\perp}^{2}-z^{2}+V^{2})\,,\label{eq:denominator-1}
\end{align}
 and $\mathcal{D}^{(4)}(z,\mathbf{\boldsymbol{\pi}})$ is the matrix\begin{widetext}

\begin{equation}
\mathcal{D}^{(4)}(z,\mathbf{\boldsymbol{\pi}})=\left(\begin{array}{cccc}
\Xi_{1+}(z,\mathbf{\boldsymbol{\pi}}) & \mathbf{\boldsymbol{\pi}}^{2}e^{2i\theta_{\mathbf{p}}}t_{\perp} & \mathbf{\boldsymbol{\pi}}e^{i\theta_{\mathbf{p}}}t_{\perp}(z+V) & \mathbf{\boldsymbol{\pi}}e^{i\theta_{\mathbf{p}}}\Upsilon_{+}(z,\boldsymbol{\pi})\\
\\
\mathbf{\boldsymbol{\pi}}^{2}e^{-2i\theta_{\mathbf{p}}}t_{\perp} & \Xi_{1-}(z,\mathbf{\boldsymbol{\pi}}) & \mathbf{\boldsymbol{\pi}}e^{-i\theta_{\mathbf{p}}}\Upsilon_{-}(z,\boldsymbol{\pi}) & \mathbf{\boldsymbol{\pi}}e^{-i\theta_{\mathbf{p}}}t_{\perp}(z-V)\\
\\
\mathbf{\boldsymbol{\pi}}e^{-i\theta_{\mathbf{p}}}t_{\perp}(z+V) & \mathbf{\boldsymbol{\pi}}e^{i\theta_{\mathbf{p}}}\Upsilon_{-}(\boldsymbol{\pi}) & \Xi_{2+}(z,\mathbf{\boldsymbol{\pi}}) & t_{\perp}(z^{2}-V^{2})\\
\\
\mathbf{\boldsymbol{\pi}}e^{-i\theta_{\mathbf{p}}}\Upsilon_{+}(z,\boldsymbol{\pi}) & t_{\perp}(z-V)\mathbf{\boldsymbol{\pi}}e^{i\theta_{\mathbf{p}}} & t_{\perp}(z^{2}-V^{2}) & \Xi_{2-}(z,\mathbf{\boldsymbol{\pi}})
\end{array}\right)\,.\label{eq:D_4}
\end{equation}
\end{widetext}In the above we have also defined the following quantities
\begin{eqnarray}
\Upsilon_{\pm}(\boldsymbol{\pi}) & = & \mathbf{\boldsymbol{\pi}}^{2}-(z\pm V)^{2}\,,\label{eq:Palmeira_pm}\\
\Xi_{1\pm}(\mathbf{\boldsymbol{\pi}}) & = & (t_{\text{\ensuremath{\perp}}}^{2}+V^{2}-z^{2})(z\pm V)+\boldsymbol{\pi}^{2}(z\mp V)\,,\label{eq:O_1pm}\\
\Xi_{2\pm}(\mathbf{\boldsymbol{\pi}}) & = & (z\pm V)\Upsilon_{\mp}(z,\boldsymbol{\pi})\,.\label{eq:O_2pm}
\end{eqnarray}
The propagator in coordinate space is obtained from (\ref{eq:propagator-G})
via inverse Fourier transform 
\begin{equation}
G(z,\mathbf{r}-\mathbf{r}^{\prime})=\int\frac{d\mathbf{p}}{4\pi^{2}}e^{i\mathbf{p}\cdot(\mathbf{r}-\mathbf{r}^{\prime})}\mathcal{G}(z,v_{F}\mathbf{p})\,.\label{eq:inv_Fourier_Transform}
\end{equation}
In order to evaluate the above integral we use differentiation under
the integral sign, that is, 
\begin{equation}
G(z,\mathbf{r}-\mathbf{r}^{\prime})=\mathcal{D}^{(4)}\left(z,\frac{\boldsymbol{\nabla}}{i}\right)\Lambda_{0}^{(4)}(z,\mathbf{r}-\mathbf{r}^{\prime})\,,\label{eq:diff_under_int_sign}
\end{equation}
with
\begin{align}
\Lambda_{0}^{(4)}(z,\mathbf{r}-\mathbf{r}^{\prime}) & =\int\frac{d\mathbf{p}}{4\pi^{2}}\frac{e^{i\mathbf{p}\cdot(\mathbf{r}-\mathbf{r}^{\prime})}}{D(z,p)}\nonumber \\
 & =\int_{0}^{\infty}\frac{dp\, p}{2\pi}\frac{J_{0}(p|\mathbf{r}-\mathbf{r}^{\prime})}{D(z,v_{F}p)}\,.\label{eq:Lambda_4}
\end{align}
 We note that the operation (\ref{eq:diff_under_int_sign}) is strictly
valid for $|\mathbf{r}-\mathbf{r}^{\prime}|\neq0$, where the inverse
Fourier transform (\ref{eq:inv_Fourier_Transform}) converges for
all matrix components of $\mathcal{G}$. The one-dimensional integral
in Eq.~(\ref{eq:Lambda_4}) can be solved using standard methods
(we provide a derivation in the Appendix~A). One obtains
\begin{align}
\Lambda_{0}^{(4)}(E,\mathbf{r})=\frac{1}{4v_{F}^{2}}\frac{1}{A_{+}+A_{-}}\times\:\:\:\:\:\:\:\:\:\:\:\:\:\:\:\:\:\:\:\:\:\:\:\:\:\:\:\:\:\:\:\:\:\:\:\:\:\:\:\:\:\:\:\:\:\:\:\:\:\:\:\:\:\:\:\:\:\nonumber \\
\begin{cases}
-iH_{0}\left(p_{+}r\right)+\frac{2}{\pi}K_{0}\left(p_{-}r\right) & \,,A_{-}>0\,(\textrm{regions II})\\
-iH_{0}\left(p_{+}r\right)+iH_{0}\left(p_{-}r\right) & \,,A_{-}<0\,(\textrm{regions I and III})
\end{cases}\,,\label{eq:Lambda_4_sol}
\end{align}
 where $p_{\pm}=\sqrt{|A_{\pm}|}/v_{F}$ and 
\begin{align}
A_{\pm} & =\pm(E^{2}+V^{2})+\sqrt{4E^{2}V^{2}+t_{\perp}^{2}(E^{2}-V^{2})}\,.\label{eq:def_A}
\end{align}
In the above, $H_{0}(.)$ {[}$K_{0}(.)${]} denotes the zeroth order
first kind Hankel {[}second kind modified Bessel{]} function. In region
II, $\Lambda_{0}^{(4)}(E,\mathbf{r})$ contains one propagating mode
$H_{0}(p_{+}r)$---representing outgoing cylindrical waves with momentum
$p_{+}$---and one evanescent wave $K_{0}(p_{-}r)$. Regions I and
III admit two (real) propagating solutions as the Fermi level intersects
the electronic bands at two distinct points {[}namely, $E=\epsilon^{\pm-}(v_{F}p_{+})$
and $E=\epsilon^{\pm-}(v_{F}p_{+})$ in region I and $E=\epsilon^{\pm-}(v_{F}p_{+})$
and $E=\epsilon^{\pm+}(v_{F}p_{-})$ in region III{]}.

Next, we show how to compute the propagator at the origin, $G(E,\mathbf{r}=0)$.
This quantity plays a central role in scattering from short-range
impurities (Sec.~\ref{sub:Lippmann-Schwinger-approach-to}). Combining
Eqs.~(\ref{eq:propagator-G})-(\ref{eq:D_4}) and (\ref{eq:inv_Fourier_Transform})
we obtain

\begin{figure}
\begin{centering}
\includegraphics[width=1\columnwidth]{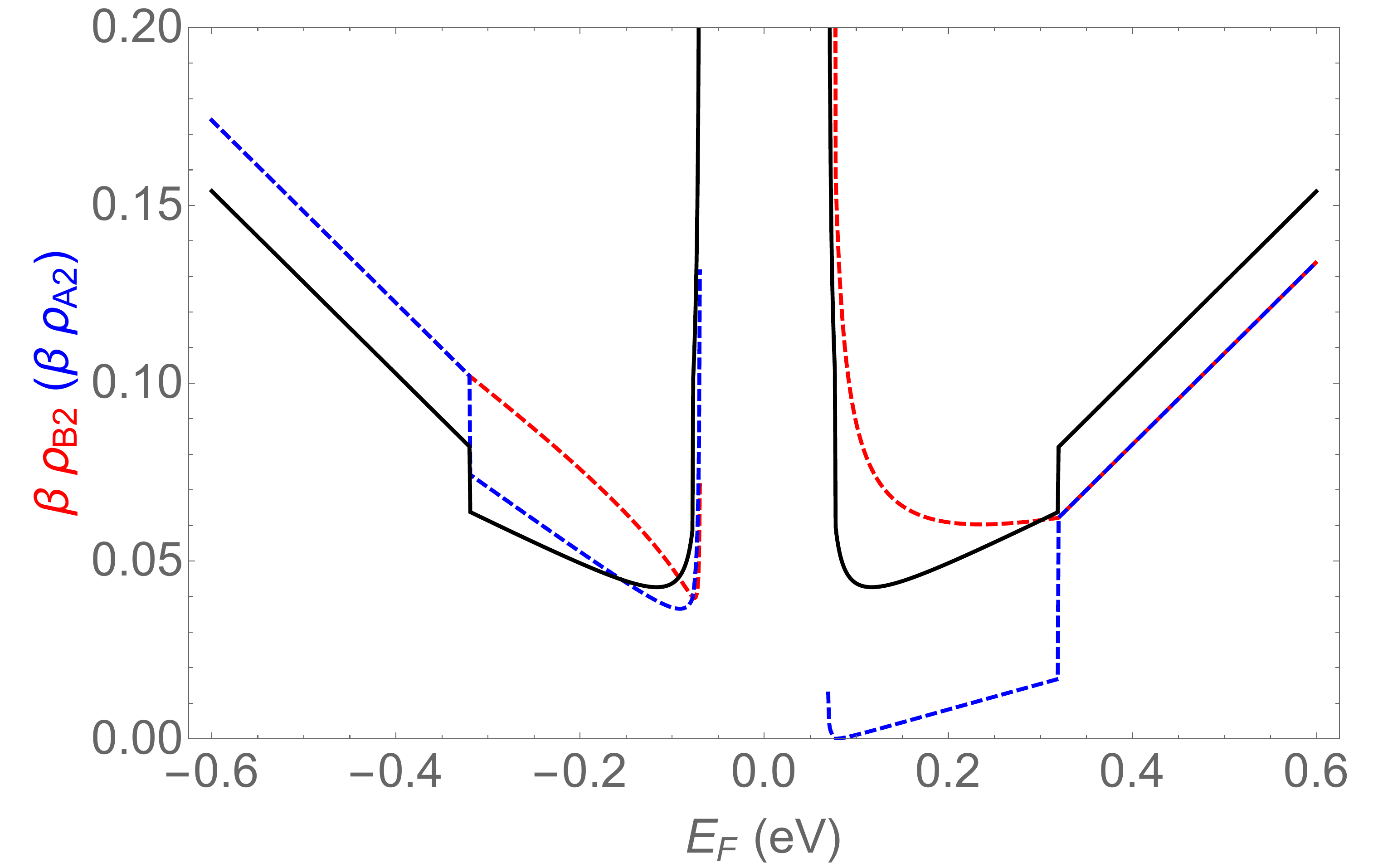} 
\par\end{centering}

\protect\protect\protect\caption{\label{fig:LDOS}The local density of states per unit area at the
bottom layer as function of energy (given in units of $\beta^{-1}=t_{\perp}/v_{F}^{2}$).
The LDOS on the top layer is obtained through reflexion, according
to $\rho_{B_{1}}(E)=\rho_{A_{2}}(-E)$ and $\rho_{A_{1}}(E)=\rho_{B_{2}}(-E)$.
The total density of states is shown in solid line. Parameters as
in Castro \emph{et al}. Ref.~\cite{BLG}: $V=77.5$ meV, $t_{\perp}=0.1t$,
and $t=3.1$ eV. }
\end{figure}

\begin{widetext} 
\begin{equation}
G(E,0)=\int_{0}^{\infty}\frac{dp}{2\pi}\frac{p}{D(E,p)+i\eta}\left[\begin{array}{cccc}
\Xi_{1+}(v_{F}p) & 0 & 0 & 0\\
0 & \Xi_{1-}(v_{F}p) & 0 & 0\\
0 & 0 & \Xi_{2+}(v_{F}p) & t_{\perp}(E^{2}-V^{2})\\
0 & 0 & t_{\perp}(E^{2}-V^{2}) & \Xi_{2-}(v_{F}p)
\end{array}\right]\,.\label{eq:G_0_int}
\end{equation}
\end{widetext}Since the matrix entries on the RHS are of the form
$a+bp^{2}$ we are lead to the following type of integrals 
\begin{eqnarray}
\Theta(E)\equiv\int_{0}^{\infty}\frac{dp}{2\pi}\frac{p}{D(E,p)+i\eta}\,,\label{eq:Integral_Regular}\\
\Theta_{\Lambda}(E)\equiv\int_{0}^{\frac{\Lambda}{v_{F}}}\frac{dp}{2\pi}\frac{p^{3}}{D(E,p)+i\eta}\,,\label{eq:Integral_Irregular}
\end{eqnarray}
with $\eta$ a real infinitesimal. We notice that the terms $bp^{2}$
result in logarithmically divergent integrals and hence a cut-off
was introduced in $\Theta_{\Lambda}(E)$. It is useful to recast the
denominator in the following form
\begin{align}
\frac{1}{D(E,p)+i\eta} & =\frac{1}{A_{-}+A_{+}}\times\nonumber \\
 & \left(\frac{1}{A_{-}+v_{F}^{2}p^{2}+is_{-}0^{+}}+\frac{1}{A_{+}-v_{F}^{2}p^{2}+i0^{+}}\right)\,,\label{eq:1/D}
\end{align}
 where $s_{-}\equiv\textrm{sign}(A_{-})$. Inserting (\ref{eq:1/D})
into (\ref{eq:Integral_Regular})-(\ref{eq:Integral_Irregular}),
and making use of the Sokhotski\textendash Plemelj formula, we arrive
at
\begin{align}
\Theta(E) & =\frac{1}{4\pi v_{F}^{2}}\frac{1}{A_{+}+A_{-}}\times\nonumber \\
 & \begin{cases}
\ln\left(\frac{A_{+}}{A_{-}}\right)-i\pi & \,,A_{-}\ge0\\
\ln\left(-\frac{A_{+}}{A_{-}}\right) & \,,A_{-}<0
\end{cases} & \,,\label{eq:Reg_Lambda_Aux}
\end{align}
\begin{align}
\Theta_{\Lambda}(E) & =-\frac{1}{4\pi v_{F}^{4}}\frac{1}{A_{+}+A_{-}}\times\nonumber \\
 & \begin{cases}
\Psi_{+}\left(\frac{\Lambda}{v_{F}}\right)+i\pi A_{+} & \,,A_{-}\ge0\\
\Psi_{-}\left(\frac{\Lambda}{v_{F}}\right)+i\pi(A_{+}+A_{-}) & \,,A_{-}<0
\end{cases} & \,,\label{eq:Irreg_Lambda_Aux}
\end{align}
where 
\begin{equation}
\Psi_{\pm}(x)=A_{-}\ln\left(\pm1+\frac{v_{F}^{2}x^{2}}{|A_{-}|}\right)+A_{+}\ln\left(-1+\frac{v_{F}^{2}x^{2}}{A_{+}}\right)\,,\label{eq:Psi}
\end{equation}
encodes the cutoff-sensitive component. The non-zero entries of Eq.~(\ref{eq:G_0_int})
become 
\begin{eqnarray}
G(E,0)_{1,1}=(t_{\text{\ensuremath{\perp}}}^{2}+V^{2}-E^{2})+(E+V)\Theta(E)\,\,\,\,\,\,\,\,\,\,\,\,\,\, &  & \notag\label{eq:G_11}\\
+\, v_{F}^{2}(E-V)\Theta_{\Lambda}(E)\,,\,\,\,\,\,\,\,\,\,\,\,\,\,\,\,\,\,\,\,\,\,\,\\
G(E,0)_{2,2}=(t_{\text{\ensuremath{\perp}}}^{2}+V^{2}-E^{2})(E-V)\Theta(E)\,\,\,\,\,\,\,\,\,\,\,\,\,\,\,\,\,\,\,\,\,\, &  & \notag\label{eq:G_22}\\
+\, v_{F}^{2}(E+V)\Theta_{\Lambda}(E)\,,\,\,\,\,\,\,\,\,\,\,\,\,\,\,\,\,\,\,\,\,\,\,\\
G(E,0)_{3,4}=t_{\perp}(E^{2}-V^{2})\Theta_{\textrm{}}(E)\,,\,\,\,\,\,\,\,\,\,\,\,\,\,\,\,\,\,\,\,\,\,\,\,\,\,\,\,\,\,\,\,\,\,\,\,\,\,\,\,\,\,\,\,\,\,\,\,\label{eq:G_34}\\
G(E,0)_{4,3}=t_{\perp}(E^{2}-V^{2})\Theta_{\textrm{}}(E)\,,\,\,\,\,\,\,\,\,\,\,\,\,\,\,\,\,\,\,\,\,\,\,\,\,\,\,\,\,\,\,\,\,\,\,\,\,\,\,\,\,\,\,\,\,\,\,\,\label{eq:G_43}\\
G(E,0)_{3,3}=(E+V)[v_{F}^{2}\Theta_{\Lambda}(E)-(E-V)^{2}\Theta(E)],\label{eq:G_33}\\
G(E,0)_{4,4}=(E-V)[v_{F}^{2}\Theta_{\Lambda}(E)-(E+V)^{2}\Theta(E)]\,.\label{eq:G_44}
\end{eqnarray}
We remark that the local density of states (LDOS) can be obtained
through a very similar calculation, according to $\rho_{ii}(E)=(-1/\pi)\textrm{Im }G^{+}(E,0)_{i,i}$
with $G^{+}(E,0)\equiv G(E+i0^{+},0)$. The different form of the
analytic continuation leads to somewhat different expressions (see
Appendix B for details). The LDOS of a representative system is shown
in Fig.~\ref{fig:LDOS}.

\subsection{Lippmann-Schwinger Approach to the Scattering Problem\label{sub:Lippmann-Schwinger-approach-to}}

We are interested in scattering potentials of the form $V_{\textrm{ad}}(\mathbf{r})=\mathcal{V}\delta(\mathbf{r})$,
where $\mathcal{V}$ is a Hermitian matrix encoding the action of
the impurity in the bilayer. The scattering problem can be solved
straightforwardly using the Lippmann-Schwinger (LS) equation \cite{Unified_MLG_BLG},
which for delta-peak potentials acquires the simple form 
\begin{equation}
\Psi_{\mathbf{k}}(\mathbf{r})=\phi_{\mathbf{k}}(\mathbf{r})+G(E,\mathbf{r})\mathcal{V}\Psi_{\mathbf{k}}(0)\,,\label{eq:LS}
\end{equation}
where $\phi_{\mathbf{k}}(\mathbf{r})$ stands for the incident wave
of energy $E$ and momentum $\mathbf{k}$, whereas $\Psi_{\mathbf{k}}(\mathbf{r})$
represents the full 'incident + scattered' wavefunction. Setting $\mathbf{r}=0$
in the LS equation results in 
\begin{equation}
\Psi_{\mathbf{k}}(0)=\left[\mathbb{I}-G(E,0)\mathcal{V}\right]^{-1}\phi_{\mathbf{k}}(0)\,,\label{eq:Psi_0}
\end{equation}
where $G(E,0)$ has been computed earlier (refer to Sec.~\ref{sec:Propagator}).
The scattered wavefunction $\Delta\Psi_{\mathbf{k}}(\mathbf{r})\equiv\Psi_{\mathbf{k}}(\mathbf{r})-\phi_{\mathbf{k}}(\mathbf{r})$
is obtained by substituting (\ref{eq:Psi_0}) into the RHS of Eq.~(\ref{eq:LS}),
i.e., 
\begin{equation}
\Delta\Psi_{\mathbf{k}}(\mathbf{r})=G(E,\mathbf{r})\mathcal{V}\left[\mathbb{I}-G(E,0)\mathcal{V}\right]^{-1}\phi_{\mathbf{k}}(0)\,.\label{eq:scatt_LS}
\end{equation}
Local electrostatic potentials are described by a diagonal matrix
in the sublattice space \cite{intervalley} 
\begin{equation}
\mathcal{V}=V_{0}\left[\begin{array}{cccc}
a & 0 & 0 & 0\\
0 & b & 0 & 0\\
0 & 0 & c & 0\\
0 & 0 & 0 & d
\end{array}\right]\,,\label{eq:V_ad}
\end{equation}
where $V_{0}\equiv Sv_{0}$ with $S$ encoding the areal range of
the potential and $v_{0}$ its strength. Denoting $g_{n_{1}n_{2}}\equiv\left[G(E,0)\right]_{n_{1},n_{2}}$
and inserting the potential matrix (\ref{eq:V_ad}) into the LS equation
(\ref{eq:scatt_LS}) we obtain 
\begin{eqnarray}
\Delta\Psi_{\mathbf{k}}(\mathbf{r})=G(E,\mathbf{r})\:\:\:\:\:\:\:\:\:\:\:\:\:\:\:\:\:\:\:\:\:\:\:\:\:\:\:\:\:\:\:\:\:\:\:\:\:\:\:\:\:\:\:\:\:\:\:\:\:\:\:\:\:\:\:\:\:\:\:\:\:\:\:\:\:\:\:\:\:\:\: &  & \notag\label{eq:scatt_sol_}\\
\left[\begin{array}{cccc}
\frac{a}{V_{0}^{-1}-ag_{11}} & 0 & 0 & 0\\
0 & \frac{b}{V_{0}^{-1}-bg_{22}} & 0 & 0\\
0 & 0 & \frac{c\left(1-dg_{44}V_{0}\right)}{\Phi(E)} & \frac{cdg_{34}V_{0}}{\Phi(E)}\\
0 & 0 & \frac{cdg_{34}V_{0}}{\Phi(E)} & \frac{d\left(1-cg_{33}V_{0}\right)}{\Phi(E)}
\end{array}\right]\phi_{\mathbf{k}}(0)\,,\label{eq:scatt_sol}
\end{eqnarray}
where 
\begin{eqnarray}
\Phi(E)=V_{0}^{-1}-dg_{44}-c\left[g_{33}+d\left(g_{34}^{2}-g_{33}g_{44}\right)V_{0}\right]\,.\label{eq:Phi}
\end{eqnarray}
Below we show how to compute the scattering flux for the particular
case: $b,c,d=0\wedge a=1$. The latter represents a potential localized
on atom $B_{2}$, for which the intervalley terms have been artificially
set to zero (see also remark \cite{intervalley}). Despite its simplicity,
this model already contains the basic physics underpinning the electron--hole
asymmetry observed in the experiments. For such a toy model, the scattered
component reads 
\begin{equation}
\left.\Delta\Psi_{\mathbf{k}}(\mathbf{r})\right|_{B_{2}}=G(E,\mathbf{r})\left[\begin{array}{cccc}
\frac{1}{V_{0}^{-1}-g_{11}} & 0 & 0 & 0\\
0 & 0 & 0 & 0\\
0 & 0 & 0 & 0\\
0 & 0 & 0 & 0
\end{array}\right]\phi_{\mathbf{k}}(0).\label{eq:Psi_scatt_}
\end{equation}

\subsection*{Solution in Region II}

To make further progress we need the form of the incident wave. Taking
the incoming momentum along the $x$ axis, and assuming for simplicity
$|V|\le|E|\le E_{H}$ (region II), we have 
\begin{align}
\left.\phi_{\mathbf{k}}(0)\right|_{\textrm{II}}=\mathcal{N}(E)\left[\begin{array}{c}
\frac{v_{F}k}{E-V}\\
\\
\frac{(E-V)(t_{\perp}^{2}+V^{2}-E^{2})+v_{F}^{2}k^{2}(E+V)}{v_{F}kt_{\perp}(E-V)}\\
\\
\frac{v_{F}^{2}k^{2}-(E-V)^{2}}{t_{\perp}(E-V)}\\
\\
1
\end{array}\right]\,,\label{eq:incident_wave}
\end{align}
with $\mathcal{N}(E)$ a normalization factor, whose form does not
need to be specified. Altogether we have 
\begin{equation}
\left.\Delta\Psi_{\mathbf{k}}(\mathbf{r})\right|_{B_{2}}=\mathcal{N}(E)G(E,\mathbf{r})\left[\begin{array}{c}
\frac{v_{F}k}{E-V}\frac{1}{V_{0}^{-1}-g_{11}(E)}\\
0\\
0\\
0
\end{array}\right].\label{eq:Psi_scatt__}
\end{equation}
All that is left is to ``propagate'' the wave by acting with $G(E,\mathbf{r})$.
The propagator reads {[}refer to Eqs.~(\ref{eq:inv_Fourier_Transform})-(\ref{eq:Lambda_4_sol}){]}
\begin{equation}
G(z,\mathbf{r})=\mathcal{D}^{(4)}\left(z,\frac{\boldsymbol{\nabla}}{i}\right)\Lambda_{0}^{(4)}(z,\mathbf{r})\,,\label{eq:prog_4}
\end{equation}
where $\mathcal{D}^{(4)}$ is obtained from Eq.~(\ref{eq:D_4}) replacing
$\mathbf{\boldsymbol{\pi}}e^{\pm i\theta_{\mathbf{p}}}\rightarrow v_{F}(\hat{p}_{x}\pm i\hat{p}_{y})$
and\begin{widetext}

\begin{align}
\left.\Lambda_{0}^{(4)}(z,\mathbf{r})\right|_{\textrm{II}} & =\frac{1}{4v_{F}^{2}}\frac{1}{A_{+}+A_{-}}\left[-iH_{0}\left(v_{F}^{-1}\sqrt{A_{+}}|\mathbf{r}|\right)+\frac{2}{\pi}K_{0}\left(v_{F}^{-1}\sqrt{|A_{-}|}|\mathbf{r}|\right)\right]\,.\label{eq:Lambda_4-1}
\end{align}
\end{widetext}Inserting this expression into (\ref{eq:prog_4}),
and using $H_{0}(x)\rightarrow\sqrt{\frac{2}{\pi x}}e^{i(x-\frac{\pi}{4})}$
for $x\gg1$, we finally arrive at

\begin{widetext} 
\begin{equation}
\left.\Delta\Psi_{\mathbf{k}}\left(r,\theta\right)\right|_{B_{2}}\overset{r\gg1}{\longrightarrow}-\frac{i}{8v_{F}^{2}}\frac{\mathcal{N}(E)}{\sqrt{4E^{2}V^{2}+t_{\perp}^{2}(E^{2}-V^{2})}}\sqrt{\frac{2}{i\pi kr}}e^{ikr}\mathcal{D}^{(4)}\left(E,v_{F}ke^{i\theta}\right)\left[\begin{array}{c}
\frac{v_{F}k}{E-V}\frac{1}{V_{0}^{-1}-g_{11}(E)}\\
0\\
0\\
0
\end{array}\right]\,,\label{eq:Psi_Scatt}
\end{equation}
\end{widetext}where we have identified the wavevector at the point
of observation {[}i.e., $\mathbf{k}^{\prime}=k\mathbf{r}/r=k(\cos\theta,\sin\theta)^{\textrm{t}}${]}.
Equation (\ref{eq:Psi_Scatt}) gives the exact asymptotic form of
the scattered wavefunctions in the energy range $|V|\le|E|\le E_{H}$.

According to the definition of differential cross section \cite{Schif_1968}
it follows that

\begin{equation}
\sigma(\theta)=r\left|\frac{v_{n}(r,\theta)}{v_{\textrm{in}}}\right|\,,\label{eq:cross_section}
\end{equation}
where $v_{\textrm{in}}$ {[}$v_{n}(r,\theta)${]} is the velocity
of the incident wave {[}scattered wave{]}, that is 
\begin{eqnarray}
v_{\textrm{in}}=v_{F}\langle\phi_{\mathbf{k}}(0)|\left(\begin{array}{cccc}
0 & 0 & 0 & 1\\
0 & 0 & 1 & 0\\
0 & 1 & 0 & 0\\
1 & 0 & 0 & 0
\end{array}\right)|\phi_{\mathbf{k}}(0)\rangle\,,\;\;\;\;\;\;\;\;\;\;\;\;\;\;\;\;\;\;\;\label{eq:v_in}\\
v_{n}(r,\theta)=v_{F}\langle\Delta\Psi_{\mathbf{k}}|\left(\begin{array}{cccc}
0 & 0 & 0 & e^{i\theta}\\
0 & 0 & e^{-i\theta} & 0\\
0 & e^{i\theta} & 0 & 0\\
e^{-i\theta} & 0 & 0 & 0
\end{array}\right)|\Delta\Psi_{\mathbf{k}}\rangle\,.\label{eq:v_in_2}
\end{eqnarray}
Substitution of (\ref{eq:Psi_Scatt}) and (\ref{eq:incident_wave})
into the above expressions and subsequent use of (\ref{eq:cross_section})
gives after somewhat lengthy but simple algebra:

\begin{widetext} 
\begin{eqnarray}
\left.\sigma(\theta)\right|_{B_{2}}^{|V|\le|E|\le E_{H}} & = & \frac{1}{32\pi k}\frac{1}{4E^{2}V^{2}+t_{\perp}^{2}(E^{2}-V^{2})}\frac{(\hbar^{2}k{}^{2}t_{\perp})^{2}}{\left|g_{11}(E)-\frac{\hbar^{2}}{SV_{0}}\right|^{2}}\frac{\Gamma(k,E,V)}{\Gamma(k,E,-V)}\,,\label{eq:sigma_A2}
\end{eqnarray}
\end{widetext}where\begin{widetext} 
\begin{equation}
\Gamma(k,E,V)=t_{\perp}^{2}v_{F}^{2}k^{2}(E+V)+\left[v_{F}^{2}k^{2}-\left(E+V\right)^{2}\right]\left[\left(E-V\right)(v_{F}k)^{2}+\left(E+V\right)\left(V^{2}+t_{\perp}^{2}-E^{2}\right)\right]\,.\label{eq:GammaEV}
\end{equation}
\end{widetext}The factor that controls the electron--hole asymmetry
is easily identified: 
\begin{equation}
\chi(E)=\frac{1}{\left|g_{11}(E)-\frac{\hbar^{2}}{SV_{0}}\right|^{2}}\frac{\Gamma(k,E,V)}{\Gamma(k,E,-V)}\,.\label{eq:chi_4}
\end{equation}
Remarkably, when the resonant scattering limit $V_{0}\rightarrow\infty$
is taken at nonzero bias $V\neq0$, the $E\rightarrow-E$ invariance
is not recovered. This is to be contrasted to the case of (i) resonant
impurities acting on both layers, and (ii) resonant impurities in
monolayer graphene, for which cross sections are electron--hole symmetric
\cite{Unified_MLG_BLG}.

\subsection*{Solution in Regions I and III}

In regions I and III, Eq.~(\ref{eq:1/D}) admits two propagating
poles because $A_{-}<0$. This occurs because the Fermi level crosses
the electronic bands at two distinct Fermi points; see Eq.~(\ref{eq:dispersion_relation})
and comments therein. As a consequence the scalar propagator {[}Eq.~(\ref{eq:Lambda_4}){]}
is given by a superposition of two cylindrical waves (see Appendix
A)\begin{widetext} 
\begin{equation}
\left.\Lambda_{0}^{(4)}(z,\mathbf{r})\right|_{|E|>E_{H}}=\frac{1}{8v_{F}^{2}}\frac{1}{\sqrt{4E^{2}V^{2}+t_{\perp}^{2}(E^{2}-V^{2})}}\left[-iH_{0}\left(k_{+}|\mathbf{r}|\right)+iH_{0}\left(k_{-}|\mathbf{r}|\right)\right]\,,\label{eq:Lambda_4_HighEnergy}
\end{equation}
\end{widetext}where 
\begin{eqnarray}
k_{\pm} & = & \frac{1}{v_{F}}\sqrt{\left(E^{2}+V^{2}\right)\pm\sqrt{4E^{2}V^{2}+t_{\perp}^{2}(E^{2}-V^{2})}}\,,\label{eq:p_F_low_high}
\end{eqnarray}
denote the Fermi momenta. The expression for the propagator evaluated
at the origin also changes because $A_{-}<0$ {[}refer to Eqs.~(\ref{eq:Reg_Lambda_Aux})-(\ref{eq:Irreg_Lambda_Aux}){]}.
The wavefunction of incident particles therefore admits two (degenerate)
solutions

\begin{equation}
\left.\phi_{\mathbf{k}}^{\pm}(0)\right|_{\textrm{I,III}}=\mathcal{N}_{\pm}(E)\left[\begin{array}{c}
\frac{v_{F}k_{\pm}}{E-V}\\
\\
\frac{(E-V)(t_{\perp}^{2}+V^{2}-E^{2})+v_{F}^{2}k_{\pm}^{2}(E+V)}{v_{F}k_{\pm}t_{\perp}(E-V)}\\
\\
\frac{v_{F}^{2}k_{\pm}^{2}-(E-V)^{2}}{t_{\perp}(E-V)}\\
\\
1
\end{array}\right]\,.\label{eq:incident_wave_HE}
\end{equation}
For instance, admitting an impurity located at site $B_{2}$ and an
incident wave with $k=k_{+}$, we have 
\begin{equation}
\left.\Delta\Psi_{\mathbf{k}}\left(r,\theta\right)\right|_{B_{2}}=\sum_{\lambda=\pm}\left.\Delta\Psi_{\mathbf{k}}^{(\lambda)}\left(r,\theta\right)\right|_{B_{2}}\,,\label{eq:psi_scatt}
\end{equation}
with\begin{widetext} 
\begin{eqnarray}
\left.\Delta\Psi_{\mathbf{k}}^{(\pm)}\left(r,\theta\right)\right|_{B_{2}} & \rightarrow & \mp\frac{i}{8v_{F}^{2}}\frac{\mathcal{N}_{\pm}(E)}{\sqrt{4E^{2}V^{2}+t_{\perp}^{2}(E^{2}-V^{2})}}\sqrt{\frac{2}{i\pi k_{\pm}r}}e^{ik_{\pm}r}\mathcal{D}^{(4)}\left(E,v_{F}k_{\pm}e^{i\theta}\right)\left[\begin{array}{c}
\frac{v_{F}k_{+}}{E-V}\frac{1}{V_{0}^{-1}-g_{11}(E)}\\
0\\
0\\
0
\end{array}\right]\,.\label{eq:psi_scatt_}
\end{eqnarray}
\end{widetext}The total cross section accounts for both the processes
$k_{+}\rightarrow k_{+}$ and $k_{+}\rightarrow k_{-}$ according
to 
\begin{align}
\left.\sigma(\theta)\right|_{k=k_{+}} & \equiv\sigma_{+}(\theta)=\sigma_{++}(\theta)+\sigma_{+-}(\theta)\,,\label{eq:diff_cross_sec}\\
\sigma_{+\pm}(\theta) & =r\frac{|v_{n}^{+\pm}(r,\theta)|}{|v_{\textrm{in}}(k_{+})|}\,,\label{eq:diff_cross_sec_2}
\end{align}
where 
\begin{equation}
v_{n}^{+\pm}(r,\theta)=v_{F}\langle\Delta\Psi_{\mathbf{k}}^{(\pm)}|\left(\begin{array}{cccc}
0 & 0 & 0 & e^{i\theta}\\
0 & 0 & e^{-i\theta} & 0\\
0 & e^{i\theta} & 0 & 0\\
e^{-i\theta} & 0 & 0 & 0
\end{array}\right)|\Delta\Psi_{\mathbf{k}}^{(\pm)}\rangle\,.\label{eq:vn_pm}
\end{equation}
Similar expressions hold for incoming waves with incoming momentum
$k_{-}$. The explicit form of $\sigma_{\pm}(\theta)$ is\begin{widetext}
\begin{equation}
\left.\sigma_{\pm}(\theta)\right|_{B_{2}}^{\textrm{I,III}}=\frac{1}{32\pi k_{\pm}}\frac{1}{4E^{2}V^{2}+t_{\perp}^{2}(E^{2}-V^{2})}\frac{(\hbar^{2}t_{\perp}k_{\pm}^{2})^{2}}{\left|g_{11}(E)-\frac{\hbar^{2}}{SV_{0}}\right|^{2}}\frac{\Gamma(k_{\pm},E,V)+\Gamma(k_{\mp},E,V)}{\Gamma(k_{\pm},E,-V)}\,.\label{eq:sigma_theta_I_III}
\end{equation}
\end{widetext}

\section{Boltzmann Transport Formulation in Biased Bilayer Graphene\label{sec:Electronic_Transport}}

\begin{figure}
\begin{centering}
\includegraphics[width=1\columnwidth]{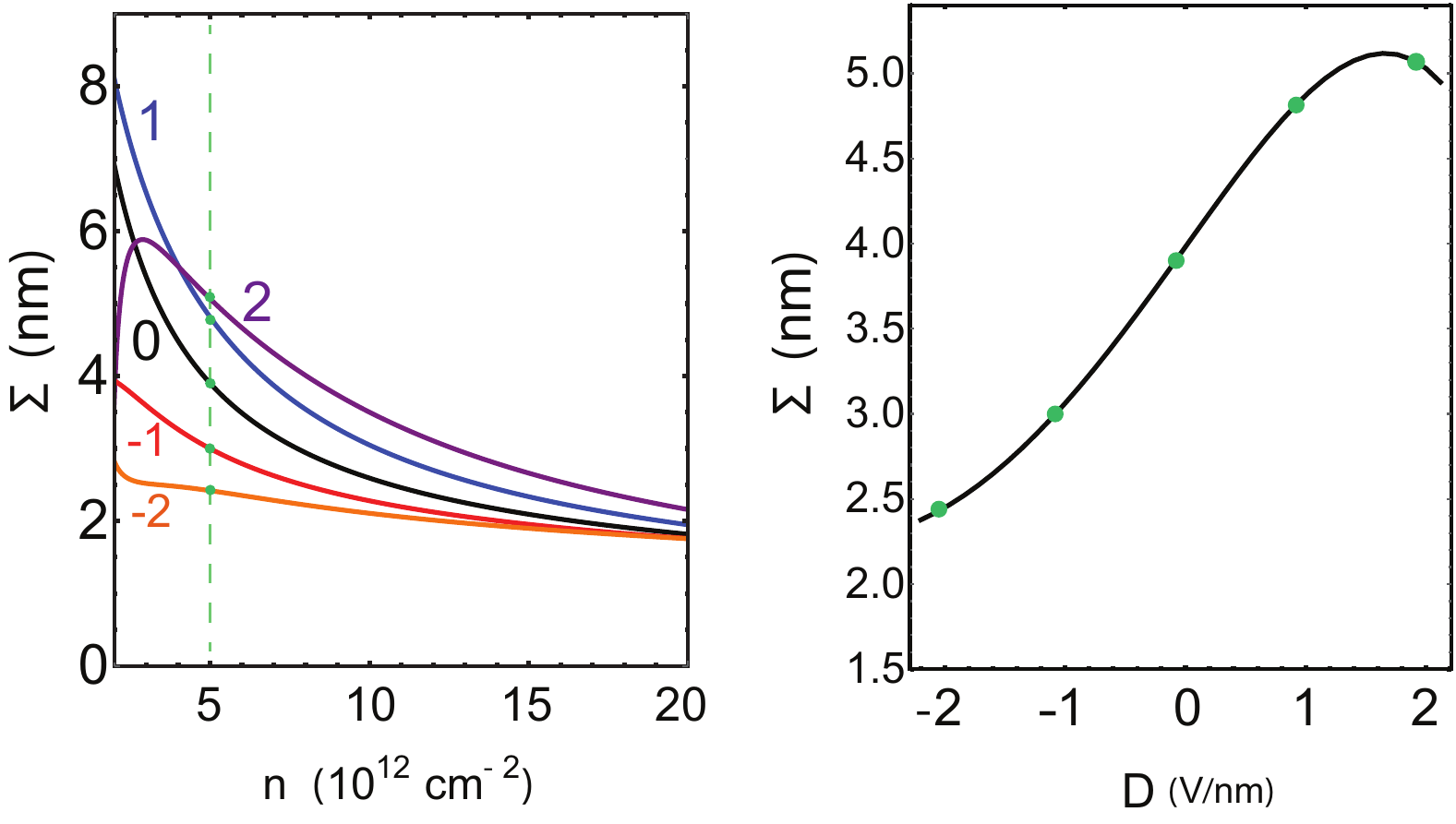} 
\par\end{centering}

\protect\protect\caption{\label{fig:CrossSection=00003D00005D}Electric field tuning of the
transport cross section. Left panel---Cross section versus electronic
density at selected values of the bias $D$-field ($D=-2\kappa V/ed$,
indicated in V/nm). Right panel---Cross section versus bias field
for $n=5\times10^{12}$ cm$^{-2}$. (Other parameters as in the main
text for sample W02.)}
\end{figure}
In this section we compute the Boltzmann dc-conductivity limited by
fluorine adatoms located in the top layer. As discussed in the main
text, fluorine adatoms act as resonant impurities with a range $R\sim0.4$~nm.
Such short-range impurities with $R\apprge a_{0}$---where $a_{0}$
denotes the carbon-carbon distance---are well described by a potential
matrix 
\begin{equation}
\mathcal{V}^{(\textrm{top})}=V_{0}\left[\begin{array}{cccc}
1 & 0 & 0 & 0\\
0 & 0 & 0 & 0\\
0 & 0 & 0 & 0\\
0 & 0 & 0 & 1
\end{array}\right]\,,\label{eq:potential_top}
\end{equation}
with the resonant limit $V_{0}\rightarrow\infty$ taken when appropriate
\cite{Unified_MLG_BLG}. The solution of the respective scattering
problem is given by Eq.~(\ref{eq:scatt_sol_}) with $a=d=1$ and
$b=c=0$. For example, in region II, one finds\begin{widetext} 
\begin{equation}
\left.\Delta\Psi_{\mathbf{k}}\left(r,\theta\right)\right|_{\textrm{top}}\overset{r\gg1}{\longrightarrow}\frac{i}{8v_{F}^{2}}\frac{\mathcal{N}(E)}{\sqrt{4E^{2}V^{2}+t_{\perp}^{2}(E^{2}-V^{2})}}\sqrt{\frac{2}{i\pi kr}}e^{ikr}\mathcal{D}^{(4)}\left(E,v_{F}ke^{i\theta}\right)\left[\begin{array}{c}
\frac{v_{F}k}{E-V}\frac{1}{g_{11}(E)}\\
0\\
0\\
\frac{1}{g_{44}(E)}
\end{array}\right]\,.\label{eq:asympt_top_region_II}
\end{equation}
\end{widetext}In the Boltzmann approach, the dc-conductivity is directly
related to the cross sections of single scattering centres (computed
in Sec.~\ref{sec:Scatt_Amp}) according to 
\begin{equation}
\textrm{\ensuremath{\sigma}}_{\textrm{dc}}(k_{F})=\frac{2e^{2}}{h}\times\begin{cases}
\frac{k_{F}}{n_{\textrm{ad}}\Sigma(k_{F})} & \,,\quad\textrm{region II}\\
\\
\sum_{s=\pm}\frac{k_{F}^{(s)}}{n_{\textrm{ad}}\Sigma(k_{F}^{(s)})} & \,,\quad\textrm{otherwise}
\end{cases}\,.\label{eq:sigma_dc-1}
\end{equation}
with $n_{\textrm{ad}}$ denoting the areal density of adatoms.The
transport cross section is given by 
\begin{eqnarray}
\Sigma(k_{F}^{(s)}) & = & \int_{0}^{2\pi}d\theta(1-\cos\theta)|\sigma_{s}(\theta)|^{2}\,,\label{eq:tau_kF_s}
\end{eqnarray}
and a similar expression holds for region II, with $\sigma(\theta)$
and $\sigma_{\pm}(\theta)$ as defined earlier. The explicit expressions
for $\Sigma(k_{F}^{(s)})$ are rather cumbersome and will not be given
here. We remark that the cross sections above are expected to match
those of a hard wall disk of radius $R\apprge a_{0}$. The equivalence
between the two scattering problems in monolayer and bilayer graphene
has been demonstrated within a two-band description by some of the
authors in Ref.~\cite{Unified_MLG_BLG}. To model the experiments
it is useful to express the Fermi energy as a function of the electronic
density\begin{widetext} 
\begin{equation}
E(n)=\textrm{sign}(n)\times\begin{cases}
\sqrt{\hbar^{2}v_{F}^{2}\pi|n|+\frac{t_{\perp}^{2}}{2}+V^{2}-\sqrt{\frac{t_{\perp}^{4}}{4}+\hbar^{2}v_{F}^{2}\pi|n|\left(t_{\perp}^{2}+4V^{2}\right)}} & \,,\quad|V|\le|E(n)|\le E_{H}\\
\sqrt{\frac{\hbar^{2}v_{F}^{2}\pi|n|}{2}-V^{2}} & \,,\quad|E(n)|>E_{H}\\
\sqrt{\frac{\left(\frac{\hbar^{2}v_{F}^{2}\pi n}{2}\right)^{2}+t_{\perp}^{2}V^{2}}{t_{\perp}^{2}+4V^{2}}} & \,,\quad\frac{\Delta}{2}<|E(n)|<|V|
\end{cases}\,.\label{eq:En_as_function_n}
\end{equation}
\end{widetext}In deriving the above formulae we have used the relations
\begin{equation}
|n|=\frac{1}{\pi}\times\begin{cases}
k_{F}^{2} & \,,|V|\le|E|\le E_{H}\\
k_{+}^{2}+k_{-}^{2} & \,,|E|>E_{H}\\
k_{+}^{2}-k_{-}^{2} & \,,\frac{\Delta}{2}<|E|<|V|
\end{cases}\,.\label{eq:n_as_function_k}
\end{equation}
These expressions were used to make the plots of the dc-conductivity
shown in the main text.

In Fig.~\ref{fig:CrossSection=00003D00005D} we show the dependence
of the cross section (\ref{eq:tau_kF_s}) on the electrostatic potential
$V$ and electronic density.\textcolor{black}{{} A few important remarks
are in order. For simplicity, we have only considered the resonant
limit $V_{0}\rightarrow\infty$. However, real fluorine induces a
finite (strong) scalar potential $V_{0}$, which breaks the exact
symmetry of our model $\sigma_{\textrm{dc}}(n,V)=\sigma_{\textrm{dc}}(-n,-V)$.
The latter and the effect of bipolar junctions (see Sec.~\ref{sec:Fabrication-and-Characterization})
could explain why experiment {[}panel (a){]} and theory {[}panel (c){]}
in Fig.~4 (main text) can only match perfectly for a given carrier
polarity. Finally, we note that the full four-band model is required
to correctly interpret the experiments, where $v_{F}p,V$ can be of
the order of $t_{\perp}$ and adatoms break the top bottom layer symmetry. }

\section{Fully Quantum Mechanical Tight-Binding Calculations using the KPM\label{sec:Fully-Quantum-Mechanical}}

\begin{center}
\begin{figure}
\begin{centering}
\includegraphics[width=1\columnwidth]{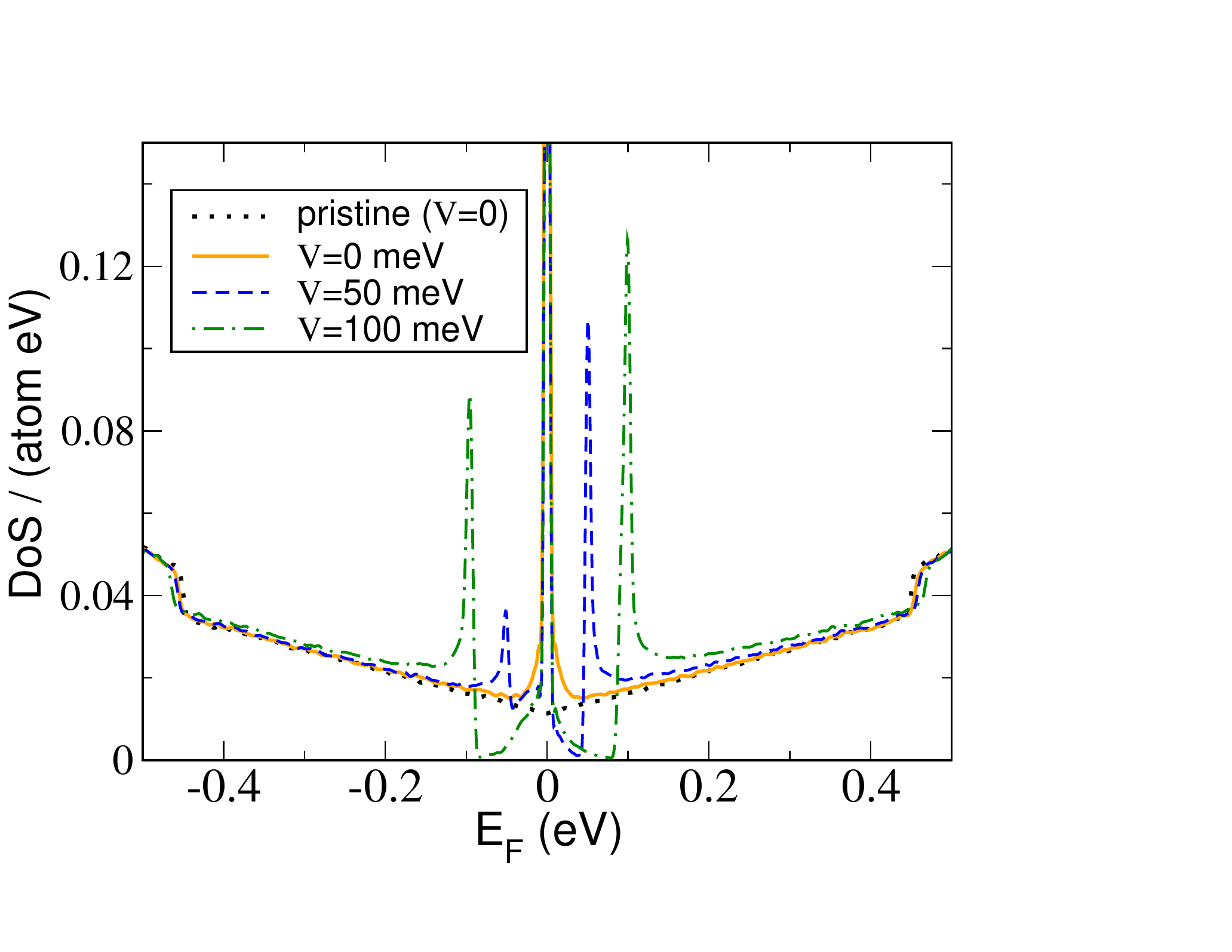} 
\par\end{centering}

\protect\caption{\label{fig:DOS_KPM}Average DOS of a real-size BLG system with $D=2\times14142^{2}$
carbon atoms at selected positive bias values for 0.05\% vacancy concentration
computed with Eq.~(\ref{eq:rho_KPM}). $N=20000$ Chebyshev moments
have been used, resulting in superior meV resolution. The DOS of pristine
unbiased system is shown as a guide to the eye. Other parameters:
$t=2.7$~eV and $t_{\perp}=0.45$~eV.}
\end{figure}

\par\end{center}

Fully quantum-mechanical calculations are performed using the kernel
polynomial method (KPM) \cite{KPM-1}. The use of KPM to compute electronic
properties of graphene was introduced in Ref.~\cite{Unified_MLG_BLG}
by some of the authors. Here, we briefly describe its application
to biased bilayer graphene (BLG) systems.

\subsubsection*{Density of States}

The density of states is formally given by 
\begin{equation}
\rho(E)=\frac{1}{D}\:\textrm{Tr}\,\delta(E-\hat{H})\,,\label{eq:rho_def}
\end{equation}
where $\hat{H}$ is tight-binding Hamiltonian \cite{TB_BLG} and $D$
is the dimension of the Hilbert space (i.e., number of lattice sites).

In the KPM approach, the---non-periodic---target function $f(x)$
is expressed in terms of a complete set of orthogonal polynomials.
First-kind Chebyshev polynomials $\{T_{n}(x)\}_{n\in\mathbb{N}_{0}}$
are usually the best choice for systems with bounded spectrum \cite{KPM,Boyd}.
The use of Chebyshev polynomials requires rescaling the spectrum into
the interval $x\in[-1:1]$, where a polynomial expansion $f(x)=\sum_{\alpha}\alpha_{n}T_{n}(x)$
is defined with the following scalar product: 
\begin{equation}
\alpha_{n}\equiv\langle T_{n}|f\rangle=\int_{-1}^{1}\frac{dx}{\pi\sqrt{1-x^{2}}}\, T_{n}(x)f(x)\,.\label{eq:KPM_}
\end{equation}
Application to the average DOS {[}Eq.~(\ref{eq:rho_def}){]} is straightforward
and results in 
\begin{equation}
\rho(\epsilon)=\frac{2}{D\pi\sqrt{1-\epsilon^{2}}}\sum_{n=0}^{\infty}\frac{1}{1+\delta_{n,0}}\,\mu_{n}T_{n}(\epsilon)\,,\label{eq:DoS_Polynom}
\end{equation}
where $\epsilon\in[-1:1]$ is the rescaled energy, and the Chebyshev
moments are given by $\mu_{n}=\textrm{Tr}\: T_{n}(\tilde{H})$, where
$\tilde{H}=\hat{H}/W$ and $W$ is the half bandwidth of the biased
bilayer system. In the KPM, the infinite sum in (\ref{eq:DoS_Polynom})
is truncated and a kernel $\{g_{n}\}_{n=0...N-1}$ is introduced to
damp Gibbs oscillations \cite{KPM-1} 
\begin{equation}
\rho_{\textrm{KPM}}(\epsilon)=\frac{2}{D\pi\sqrt{1-\epsilon^{2}}}\sum_{n=0}^{N-1}\frac{g_{n}}{1+\delta_{n,0}}\,\mu_{n}T_{n}(\epsilon)\,,\label{eq:rho_KPM}
\end{equation}
where $N$ depends on the desired resolution $\eta$, usually $N\propto W/\eta$.
For the DOS we use the Jackson kernel due to its superior performance
close to the Dirac point \cite{Unified_MLG_BLG} 
\begin{equation}
g_{n}=\frac{\left(N-n+1\right)\cos\left(\frac{\pi n}{N+1}\right)+\cot\left(\frac{\pi}{N+1}\right)\sin\left(\frac{\pi n}{N+1}\right)}{N+1}\,.\label{eq:Jackson_Kernel}
\end{equation}

Finally, to compute the moments $\mu_{n}$ efficiently we make use
of the stochastic trace evaluation (STE) technique \cite{Ebisuzaki_2004}.
For large sparse matrices, the STE amounts to replace the trace $\textrm{Tr}$
in Eq.~(\ref{eq:rho_def}) by the average with respect to a single
random vector $|R\rangle=\sum_{i=1...D}\,\chi_{i}|i\rangle$, that
is, $\textrm{Tr}\: T_{n}(\hat{H})\rightarrow\langle R|T_{n}(\hat{H})|R\rangle$.
The latter is essential exact for large sparse systems because fluctuations
can be shown to be of the order of $D^{-1/2}$.

As explained in the main text, we model resonant adatoms by removing
the corresponding $p_{z}$ orbitals to which they hybridize \cite{Unified_MLG_BLG}.
The average DOS of real-size bilayer graphene system with dilute vacancies
in the top layer computed with this powerful technique is shown in
Fig.~\ref{fig:DOS_KPM} for several values of the bias inter-layer
potential.

\begin{figure}
\begin{centering}
\includegraphics[width=0.9\columnwidth]{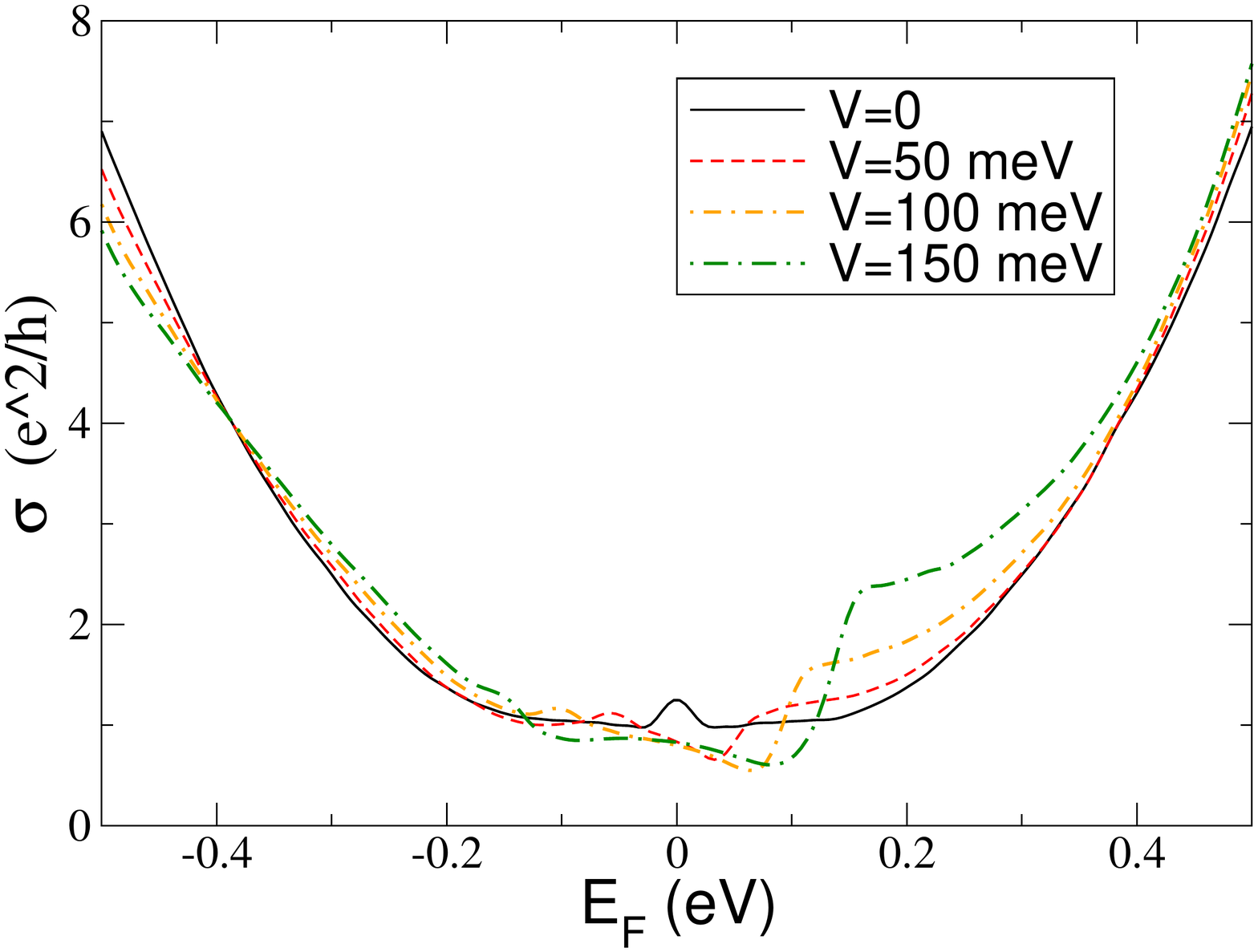}
\par\end{centering}

\begin{centering}
\includegraphics[width=0.9\columnwidth]{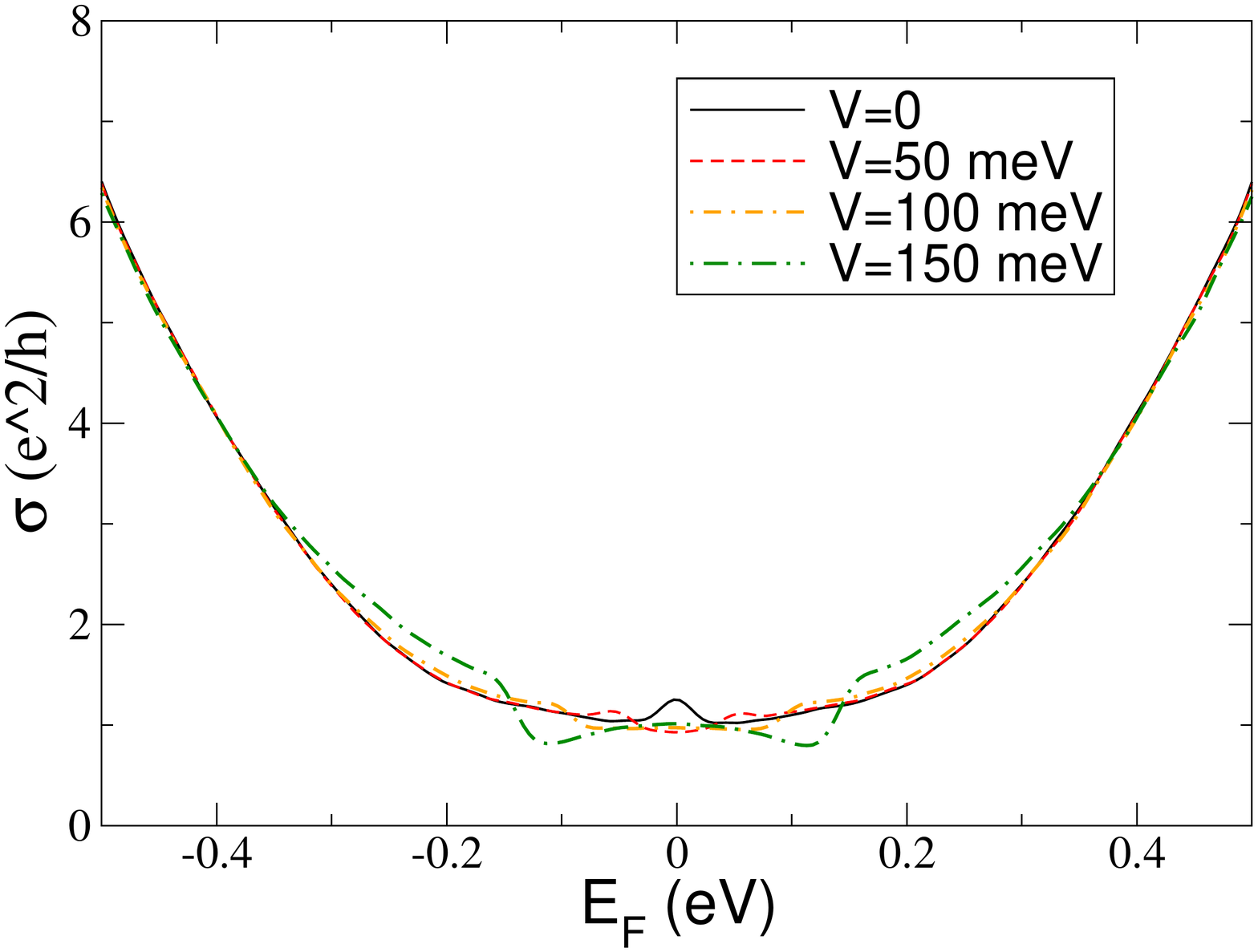} 
\par\end{centering}

\protect\protect\protect\protect\caption{\label{fig:COND_KPM}Kubo dc-conductivity of a BLG system with $D=2\times3200^{2}$
carbon atoms at selected bias values for 0.5\% vacancy concentration
{[}(top panel) vacancies distributed in the right layer; (bottom panel)
vacancies distributed in the both layers{]}. The calculation has $N^{2}=1000^{2}$
Chebyshev expansion coefficients and kernel resolution $\eta=15$
meV. Other parameters as in Fig.~\ref{fig:DOS_KPM}.}
\end{figure}

\subsubsection*{DC-Conductivity}

Next we discuss the application of KPM to the calculation of the dc-conductivity.
The starting point is the Kubo formula at zero temperature \cite{Mahan}
\begin{equation}
\sigma_{\textrm{dc}}(E)=\frac{\pi g_{s}\hbar e^{2}}{\Omega}\textrm{Tr}\left[\hat{v}_{x}\,\delta(E-\hat{H})\,\hat{v}_{x}\,\delta(E-\hat{H})\right],\label{eq:Kubo_formula}
\end{equation}
where $\hat{v}_{x}=(i/\hbar)[\hat{H},\hat{x}]$ is the $x$-th component
of the velocity operator, $g_{s}=2$ is a spin degeneracy factor,
and $\Omega$ stands for the area. In order to write $\sigma_{\textrm{dc}}(E)$
in terms of Chebyshev polynomials, we rescale the energy variable
and operators in (\ref{eq:Kubo_formula}) to get 
\begin{equation}
\sigma_{\textrm{dc}}(\epsilon)=\frac{\pi g_{s}\hbar e^{2}}{\Omega}\textrm{Tr}\left[\tilde{v}_{x}\,\delta(\epsilon-\tilde{H})\,\tilde{v}_{x}\,\delta(\epsilon-\tilde{H})\right]\,.\label{eq:Kubo_rescaled}
\end{equation}
The above formula has support in the interval $[-1:1]$ and can now
be written in terms of a KPM expansion as 
\begin{equation}
\sigma_{\textrm{KPM}}(\epsilon)=\frac{2g_{s}\hbar e^{2}}{\pi\Omega\left(1-\epsilon^{2}\right)}\sum_{n,m=0}^{N-1}\Delta_{nm}\,\left[T_{n+m}(\epsilon)+T_{|n-m|}(\epsilon)\right]\,,\label{eq:sigma_KPM}
\end{equation}
where the expansion coefficients 
\begin{equation}
\Delta_{nm}\equiv\frac{g_{n}}{1+\delta_{n,0}}\frac{g_{m}}{1+\delta_{m,0}}\textrm{Tr}\left[\tilde{v}_{x}\, T_{n}(\tilde{H})\,\tilde{v}_{x}\, T_{m}(\tilde{H})\right]\,,\label{eq:Delta_nm}
\end{equation}
are computed with the STE technique, and a Lorentz kernel \cite{KPM-1}
\begin{equation}
g_{n}=\frac{\sinh[\tilde{\eta}(N-n)]}{\sinh(\tilde{\eta}N)}\,,\label{eq:Lorentzian_Kernel}
\end{equation}
is used to approximate the delta functions appearing in Eq.~(\ref{eq:Kubo_rescaled})
by Lorentzians with resolution $\eta\equiv\tilde{\eta}W$. The computational
cost of evaluating the matrix expansion coefficients in Eq.~(\ref{eq:Delta_nm})
is now substantially higher than computing DOS moments $\mu_{n}=\textrm{Tr}\: T_{n}(\tilde{H})$,
which in practice limits the size of systems we can tackle.

In Fig.~\ref{fig:COND_KPM} we show the Kubo dc-conductivity for
a BLG system with $\sim$20 million atoms %
\footnote{In order to reach the diffusive regime we have artificially increased
the concentration of impurities as compared to the experimental determined
values. %
}. The results disclose a conductivity plateau in the pseudo-gap region
due to an impurity band formation, as found earlier in Ref.~\cite{Unified_MLG_BLG}.
More importantly, the bias potential is seen to originate electron--hole
assymetry only for impurities restricted to a single layer, as discussed
in the main text. Remarkably---away from the Dirac point---the conductivity
displays the exact same trends as predicted by the semi-classical
calculations. In particular, the linear response calculation shows
that the increase of the conductivity upon increasing the bias field
saturates in the electron sector ($E_{F}>0$), exactly as predicted
by the 4-band semi-classical model (Sec.~\ref{sec:Electronic_Transport}).
The opposite trends are observed upon reversal of the bias sign (not
shown), again as predicted by the semi-classical calculation. \pagebreak{}

\section{Appendix A: Calculation of the Propagator in Coordinate Space\label{sec:Appendix_I}}

For completeness we show how to compute the integral 
\begin{equation}
\Lambda_{0}^{(4)}(\mathbf{u})=\int_{0}^{\infty}\frac{dp\, p}{2\pi}\frac{J_{0}(pu)}{D(z,v_{F}p)+i\eta}\,,\quad u\equiv|\mathbf{u}|>0\,.\label{eq:int_-1}
\end{equation}
The sign of the infinitesimal $\eta$ is chosen as to guarantee that
$\Lambda_{0}^{(4)}(\mathbf{u})$ represents an outgoing wave. Using
Eq.~(\ref{eq:1/D}) we split the integral into two contributions
$\Lambda_{0}^{(4)}(\mathbf{u})=\Lambda_{0}^{+}(\mathbf{u})+\Lambda_{0}^{-}(\mathbf{u})$,
where 
\begin{equation}
\Lambda_{0}^{\pm}(\mathbf{u})=\int_{0}^{\infty}\frac{dp\, p}{2\pi a}\frac{J_{0}(pu)}{A_{\pm}\mp v_{F}^{2}p^{2}+i0^{+}}\,,\label{eq:Lambda_p_m}
\end{equation}
and $a=A_{-}+A_{+}$ is a constant. The cases $A_{-}\gtrless0$ have
to be considered separately. When $A_{-}>0$ (region II), the term
$\Lambda_{0}^{-}(\mathbf{u})$ reads 
\begin{equation}
\Lambda_{0}^{-}(\mathbf{u})=\int_{0}^{\infty}\frac{dp\, p}{2\pi a}\frac{J_{0}(pu)}{A_{-}+v_{F}^{2}p^{2}+i0^{+}}=\frac{1}{a}K_{0}(\sqrt{A_{-}}u)\,.\label{eq:Lambda_-}
\end{equation}
In the above we have made use of 
\[
\int_{0}^{\infty}dp\, p\frac{J_{0}(pu)}{p^{2}+z^{2}}=K_{0}\left(zu\right)\quad,\,\textrm{for}\;\textrm{Re}\, z>0\,,
\]
(see, e.g., Eq.~EH II 96(58), Ref.~\cite{Gradshteyn}), whereas
for $A_{-}<0$ (regions I and III) one finds \cite{analytic_cont}
\begin{equation}
\Lambda_{0}^{-}(u)=\int_{0}^{\infty}\frac{dp\, p}{2\pi a}\frac{J_{0}(pu)}{-|A_{-}|+v_{F}^{2}p^{2}-i0^{+}}=\frac{i\pi}{2a}H_{0}(\sqrt{|A_{-}|}u).\label{eq:Lambda_m_2}
\end{equation}
The term $\Lambda_{0}^{+}(\mathbf{u})$ has the same form in all regions
since $A_{+}>0$. We find
\begin{equation}
\Lambda_{0}^{+}(u)=\int_{0}^{\infty}\frac{dp\, p}{2\pi a}\frac{J_{0}(pu)}{A_{+}-v_{F}^{2}p^{2}+i0^{+}}=-\frac{i\pi}{2a}H_{0}(\sqrt{A_{+}}u).\label{eq:Lambda_p}
\end{equation}
The meaning of all these terms is clear. Eqs.~(\ref{eq:Lambda_m_2})-(\ref{eq:Lambda_p})
represent propagating waves with the characteristic two-dimensional
asymptotic behavior $\propto e^{iku}/\sqrt{ku}$, whereas (\ref{eq:Lambda_-})
encodes an evanescent mode.

\section{Appendix B: Calculation of the Pristine LDOS\label{sec:Appendix_II}}

The local density of states (LDOS) is formally given by 
\begin{equation}
\rho(E)=-\frac{1}{\pi}\left.\textrm{Im }\textrm{diag}\left\{ G(z,0)_{1,1},\:...\:,G(z,0)_{4,4}\right\} \right|_{z\rightarrow E+i0^{+}}\,.\label{eq:LDOS_def}
\end{equation}
From Eq.~(\ref{eq:propagator-G}) we obtain the LDOS as an integral
over the momentum variable
\begin{align}
\rho(E) & =-\frac{1}{\pi}\;\textrm{Im }\int_{0}^{\infty}\frac{dp}{2\pi}\frac{p}{\prod_{\lambda,\lambda^{\prime}=\pm1}\left[E-\epsilon^{\lambda\lambda^{\prime}}(v_{F}p)+i0^{+}\right]}\times\nonumber \\
 & \qquad\textrm{diag}\left\{ \Xi_{1+}(v_{F}p),\Xi_{1-}(v_{F}p),\Xi_{2+}(v_{F}p),\Xi_{2-}(v_{F}p)\right\} \,.\label{eq:aux_ldos-1}
\end{align}
 Using the Sokhotski\textendash Plemelj formula and evaluating the
imaginary part we get after straighfoward integrations 
\begin{eqnarray}
\rho(E)_{1,1} & = & (t_{\text{\ensuremath{\perp}}}^{2}+V^{2}-E^{2})(E+V)\varPhi^{+}+(E-V)\varPhi_{\Lambda}^{+}\,,\label{eq:G_11-1}\\
\rho(E)_{2,2} & = & (t_{\text{\ensuremath{\perp}}}^{2}+V^{2}-E^{2})(E-V)\varPhi^{+}+(E+V)\varPhi_{\Lambda}^{+}\,,\label{eq:G_22-1}\\
\rho(E)_{3,3} & = & (E+V)\left[\varPhi_{\Lambda}^{+}-(E-V)^{2}\varPhi^{+}\right]\,,\label{eq:G_33-1}\\
\rho(E)_{4,4} & = & (E-V)\left[\varPhi_{\Lambda}^{+}-(E+V)^{2}\varPhi^{+}\right]\,,\label{eq:G_44-1}
\end{eqnarray}
with\begin{widetext} 
\begin{eqnarray}
\left[\begin{array}{c}
\varPhi_{\textrm{}}^{+}\\
\varPhi_{\textrm{\ensuremath{\Lambda}}}^{+}
\end{array}\right] & = & \frac{s_{E}}{4\pi v_{F}^{2}}\frac{1}{A_{+}+A_{-}}\times\begin{cases}
\left[\begin{array}{c}
1\\
A_{+}
\end{array}\right] & \,,\quad|V|\le|E|\le E_{H}\\
\\
\left[\begin{array}{c}
2\\
A_{+}-A_{-}
\end{array}\right] & \,,\quad\frac{\Delta}{2}<|E|<|V|\\
\\
\left[\begin{array}{c}
0\\
A_{+}+A_{-}
\end{array}\right] & \,,\quad|E|>E_{H}
\end{cases}\,.\label{eq:Reg_Lambda_Aux-1-1}
\end{eqnarray}
\end{widetext}\begin{widetext}Explicitly, 
\[
\left[\begin{array}{c}
\varPhi_{\textrm{}}^{+}\\
\varPhi_{\textrm{\ensuremath{\Lambda}}}^{+}
\end{array}\right]=\frac{s_{E}}{8\pi v_{F}^{2}}\frac{1}{\sqrt{4E^{2}V^{2}+t_{\perp}^{2}(E^{2}-V^{2})}}\times\begin{cases}
\left[\begin{array}{c}
1\\
E^{2}+V^{2}+\sqrt{4E^{2}V^{2}+t_{\perp}^{2}(E^{2}-V^{2})}
\end{array}\right] & \,,\quad|V|\le|E|\le E_{H}\\
\\
\left[\begin{array}{c}
2\\
2\left(E^{2}+V^{2}\right)
\end{array}\right] & \,,\quad\frac{\Delta}{2}<|E|<|V|\\
\\
\left[\begin{array}{c}
0\\
2\sqrt{4E^{2}V^{2}+t_{\perp}^{2}(E^{2}-V^{2})}
\end{array}\right] & \,,\quad|E|>E_{H}
\end{cases}\,.
\]
\end{widetext}


\begin{thebibliography}{10}
\bibitem{screening}A weak $n$-dependence of screening $\kappa(n)$
arising from charge imbalance between the layers has been neglected.
\end{thebibliography}
\begin{thebibliography}{10}
\bibitem{intervalley} Local potentials with range $R\apprle a$---where
$a$ denotes the lattice spacing---induce intervalley scattering and
thus a rigorous scattering treatment requires dealing with the valley
subspace explicitly. We note, however, that our interest is on adatoms
having $R\gtrsim a$. For such potentials the current formalism is
adequate provided the correct ansatz for $\mathcal{V}$ is taken (see
Sec.~\ref{sec:Electronic_Transport} and Ref.~\cite{Unified_MLG_BLG}).
Furthermore, the Kubo conductivity within a tight-binding formulation
(Sec.~\ref{sec:Fully-Quantum-Mechanical}) is seen display the same
trends observed in the semiclassical calculation, hence providing
further support to our minimal low-energy model.
\bibitem{TB_BLG}We use the following tight-binding model 
\begin{align*}
\hat{H}_{\textrm{BLG}} & =\sum_{\langle i,j\rangle}\left(-ta_{i,1}^{\dagger}b_{j,1}-ta_{i,2}^{\dagger}b_{j,2}-t_{\perp}a_{i,1}^{\dagger}b_{j,2}+\textrm{H.c.}\right)\\
 & +V\sum_{i}\sum_{L=1,2}(-1)^{L}\left(a_{i,L}^{\dagger}a_{i,L}+b_{i,L}^{\dagger}b_{i,L}\right)\,,
\end{align*}
where $a(b)_{i,L}^{\dagger}$ creates quasiparticles in site $i$
in sublattice $A(B)$ at bottom/top layer $L=1,2$, $t(t_{\perp})$
is the $A$-$B$ intra(inter)-layer hopping integral, and $2V$ is
the potential difference across the layers. To distinguish electron--hole
asymmetry induced by defects from band structure effects, intra- and
inter-layer second-next-neighbor hoppings have been neglected \cite{fine_features}.
\end{thebibliography}
\end{document}